\documentclass[preprint,superscriptaddress,nofootinbib]{revtex4-1}
\usepackage{graphicx}
\usepackage{dcolumn}
\usepackage{bm}
\usepackage{amsmath}
\usepackage{amsfonts} 
\usepackage{latexsym}
\usepackage{bbm}
\usepackage{color}
\usepackage{amssymb}
\usepackage{amsthm}
\usepackage{epsf}
\usepackage{epsfig}
\usepackage{caption3}
\usepackage{appendix}
\usepackage{cases}
\usepackage{subfigure}
\usepackage{multirow}
\usepackage[colorlinks,linkcolor=blue,anchorcolor=blue,citecolor=blue,urlcolor=blue,]{hyperref}
\makeatletter

\newcommand{\Rmnum}[1]{\expandafter\@slowromancap\romannumeral #1@}
\makeatother

\begin{document}

\title{Energy budget and the gravitational wave spectra beyond the bag model}

\author{Xiao Wang}%
\email{wangxiao2016@ihep.ac.cn}
\affiliation{Theoretical Physics Division, Institute of High Energy Physics, Chinese Academy of Sciences, 19B Yuquan Road, Shijingshan District, Beijing 100049, China}
\affiliation{School of Physics, University of Chinese Academy of Sciences, Beijing 100049, China}

\author{Fa Peng Huang}%
\email{Corresponding author. huangfp8@sysu.edu.cn}
\affiliation{Department of Physics and McDonnell Center for the Space Sciences, Washington University, St.
	Louis, Missouri 63130, USA}
\affiliation{TianQin Research Center for Gravitational Physics and School of Physics and Astronomy,
Sun Yat-sen University (Zhuhai Campus), Zhuhai 519082, China}

\author{Xinmin Zhang}
\affiliation{Theoretical Physics Division, Institute of High Energy Physics, Chinese Academy of Sciences, 19B Yuquan Road, Shijingshan District, Beijing 100049, China}
\affiliation{School of Physics, University of Chinese Academy of Sciences, Beijing 100049, China}

\bigskip


\begin{abstract}
The energy budget of cosmological first-order phase transition is essential for the gravitational wave spectra. 
Most of the previous studies are based on the bag model with same sound velocity in the symmetric and broken phase.
We study the energy budget and the corresponding gravitational wave spectra beyond the bag model, where the sound velocities could be different in the symmetric and broken phase. 
Taking the Higgs sextic  effective model as a representative model, we calculate the sound velocities in different phase, the gravitational wave spectra and the signal-to-noise ratio for different combinations of phase transition parameters beyond the bag model. We compare these new results with the ones obtained from the bag model.
The proper sound velocities and  phase transition parameters at the appropriate temperature are important to obtain more precise predictions.
\end{abstract}


\maketitle

\section{Introduction}
First-order phase transition (FOPT)  in the early Universe might play important roles in the physical processes of  baryogenesis, 
dark matter, primordial magnetic field, primordial black hole, 
cosmic string, gravitational wave (GW), etc.
The GWs generated during a strong FOPT through bubble collision, turbulence and sound wave mechanisms could provide new signals to unravel the above problems, and can be potentially detected by the future GW experiments, such as 
LISA~\cite{Audley:2017drz,Caprini:2015zlo,Caprini:2019egz,LISA:documents},
TianQin~\cite{Luo:2015ght,Hu:2018yqb,Mei:2020lrl}, Taiji~\cite{Hu:2017mde,Guo:2018npi}, Decihertz Interferometer Gravitational Wave Observatory (DECIGO)~\cite{Seto:2001qf,Kawamura:2011zz},
Ultimate-DECIGO(U-DECIGO)~\cite{Kudoh:2005as},
and Big Bang Observer (BBO)~\cite{Corbin:2005ny}.
Therefore, to clearly understand these fundamental issues in particle physics and cosmology, it is crucial to precisely predict the GW spectra produced by the FOPT process.
And the energy budget of cosmological FOPT is essential for precise calculations of GW spectra~\cite{KurkiSuonio:1984ba,KurkiSuonio:1995pp,Kamionkowski:1993fg,Espinosa:2010hh,Leitao:2010yw,Leitao:2014pda,Giese:2020rtr,Hindmarsh:2019phv}.
To obtain the energy budget, it is key to calculate the kinetic energy fraction, which is determined by hydrodynamics of expanding bubble, the phase transition strength, the sound velocity in the plasma and the bubble wall velocity.
For simplicity, most of previous studies use the bag model \cite{KurkiSuonio:1984ba,KurkiSuonio:1995pp,Kamionkowski:1993fg,Espinosa:2010hh,Leitao:2010yw,Hindmarsh:2019phv}, which assumes the symmetric and broken phase share the same constant sound velocity $c_s^2=1/3$, to describe the phase transition.
However, the phase transition process of a given new physics model could deviate from the bag model, if some extra particle contents obtain field-dependent masses that are comparable with the phase transition temperature,
and this situation could be general in many extensions of the standard model of particle physics.
Therefore, some recent studies consider a situation beyond the bag model~\cite{Leitao:2014pda,Giese:2020rtr}.
In those works, the symmetric and broken phase can be quantified by a phenomenological equation of state (EOS), and the sound velocities in the two phases could be different.
In this work, we study the different sound velocities model (DSVM) of EOS, which is one simple and natural generalization of the bag model, to explore more reliable energy budget and hydrodynamical processes of this EOS,
and the corresponding profiles of the DSVM for different hydrodynamical modes are obtained by solving the fluid equations with different sound velocities in the symmetric and broken phase.
Taking the Higgs sextic effective model as an example, we consider the effect of different sound velocities in broken and symmetric phase.  
Then, based on this effective model we show a concrete calculation of sound velocity, the kinetic energy fraction and other phase transition parameters.
For different combination of these phase transition parameters, which are obtained by the bag model of EOS and the DSVM of EOS at different characteristic temperatures, we calculate the GW spectra and observe differences of different parameter combinations.
And according to these GW spectra, we also compare signal-to-noise ratio (SNR) of different combinations of phase transition parameters for different models of EOS, which are essential for a more reliable prediction. 

This paper is organized as the following:
In Sec.~\ref{hd}, we discuss the hydrodynamics of the plasma and the corresponding EOS to quantify it.
Then, we discuss how to calculate the kinetic energy fraction in Sec.~\ref{kef}.
In Sec.~\ref{md}, we study the Higgs sextic model to perform the concrete calculations,
and the corresponding GW spectra and signal-to-noise ratio (SNR) of different parameter combinations are investigated in Sec.~\ref{gwsnr}.
Discussions and conclusions are given in Sec.~\ref{dis} and Sec.~\ref{cn}.
The Appendixes give fluid profiles of three hydrodynamical modes and the efficiency parameter.

\section{The Hydrodynamics and the Equation of State}\label{hd}
To calculate the kinetic energy fraction, we should firstly solve the hydrodynamical equations to get the 
the fluid profile.
Assuming the plasma as ideal fluid, 
we begin our discussions from the energy-momentum  
tensor of the perfect fluid
\begin{equation}
T^{\mu\nu} = (e+p)u^\mu u^\nu - pg^{\mu\nu}= w u^\mu u^\nu - pg^{\mu\nu}
\end{equation}
where $p$ and $e$ are the  pressure and energy density, 
and $u^{\mu }=\gamma (1,\mathbf{v})$
with $\gamma =1/\sqrt{1-v^{2}}$ is the four-velocity. 
 The enthalpy $w=e+p$.
These quantities can be obtained from the equilibrium free energy
density $\mathcal{F}$ in each phase, which is just the finite-temperature effective potential.
The pressure is obtained by $p=-\mathcal{F}$.
Further, we can obtain the entropy density by $s=\partial p/\partial T$, the energy density by $e=Ts-p$, and the enthalpy by $w=e+p=Ts$.

From the energy-momentum conservation $\partial_\mu T^{\mu\nu} = 0$, we can obtain the matching conditions (in the reference frame of the bubble wall) \cite{KurkiSuonio:1984ba,Kamionkowski:1993fg,KurkiSuonio:1995pp,Espinosa:2010hh,Leitao:2010yw}
\begin{equation}
w_-v_-^2\gamma_-^2 + p_- = w_+v_+^2\gamma_+^2 + p_+,\quad w_-v_-\gamma_-^2 = w_+v_+\gamma_+^2\,\,.
\end{equation}
Equivalently, we can obtain the following relations:
\begin{equation}
v_+v_- = \frac{p_+ - p_-}{e_+ - e_-},\quad \frac{v_+}{v_-} = \frac{e_- + p_+}{e_+ + p_-}\,\,.
\end{equation}
Here subscripts + and $-$ indicate the quantities in the symmetric and broken phase, respectively.
In the literature, one usually assumes the bag model of EOS to continue the analysis.
For the bag model, the EOS can be written as
\begin{equation}
p_+ = \frac{1}{3}a_+T_+^4 - \epsilon_+,\quad e_+ = a_+T_+^4 + \epsilon_+\,\,,\notag
\end{equation}
\begin{equation}
p_- = \frac{1}{3}a_-T_-^4 - \epsilon_-,\quad e_- = a_-T_-^4 + \epsilon_- \,\,,
\end{equation}
and this introduces the conventional definition of the phase transition strength parameter~\cite{Espinosa:2010hh}
\begin{equation}
\alpha_{\theta} = \frac{4}{3}\frac{\Delta\epsilon}{w_+},\quad \epsilon_{\pm} = \frac{1}{4}(e_{\pm} - 3p_{\pm})\,\,.
\end{equation}
We use natural units with $c=\hbar=k_B=1$. 
In the bag model, the sound velocity of symmetric phase and broken phase both equal $1/\sqrt{3}$.
However, the sound velocity can basically deviate from this value in a realistic FOPT process.
In general, the sound velocity should also be temperature-dependent.

To more precisely describe the phase transition process, we can assume the sound velocities in both phases are constants that can deviate from $1/\sqrt{3}$,
and we can have a relation as $\partial p/\partial e = c_s^2 = \rm constant$.
Hence the bag model of EOS can be generalized as a DSVM of EOS (which is firstly studied by Ref.~\cite{Leitao:2014pda} with the planar approximation, and also called the $\nu$ model \cite{Giese:2020rtr}).
The generalized EOS is 
\begin{equation}
p_+ = c_+^2a_+T_+^{\nu_+} - \epsilon_+,\quad e_+ = a_+T_+^{\nu_+} + \epsilon_+\,\,,\notag
\end{equation}
\begin{equation}
p_- = c_-^2a_-T_-^{\nu_-} - \epsilon_-,\quad e_- = a_-T_-^{\nu_-} + \epsilon_-\,\,,
\end{equation}
where $\nu_{\pm} = 1 + 1/c_{\pm}^2$. 
For $c_-^2 = c_+^2 = 1/3$ (i.e. the bag model), $a_-$ and $a_+$ are dimensionless and well defined.
However, for $c_{\pm}^2 \ne 1/3$, the coefficient $a_{\pm}$ is dimensional.
Since the bag model of EOS is a special situation of this general model, the temperature-dependent coefficient $a_{\pm}(T)$ can be defined as $a_{\pm}(T) = a_{\pm}T^{4-\nu_{\pm}}$, where the dimensionless constant $a_{\pm}$ is the same as the bag model of EOS. 
Hence when $c_{\pm}^2 = 1/3$, this model can return to the bag model of EOS,
and this model can be rewritten as follows:
\begin{equation}
p_+ = c_+^2a_+T_+^4 - \epsilon_+,\quad e_+ = a_+T_+^4 + \epsilon_+\,\,,\notag
\end{equation}
\begin{equation}
p_- = c_-^2a_-T_-^4 - \epsilon_-,\quad e_- = a_-T_-^4 + \epsilon_-\,\,,\label{constantmodel}
\end{equation}
with $a_{\pm} = g_{\pm}\pi^2/30$ ($g_{\pm}$ is degree of freedom for the symmetric and broken phase) just like the bag model of EOS.

The strength parameter defined in the bag model actually depends on $T_+$ and $T_-$.
That is, $\Delta\epsilon = \epsilon_+(T_+) - \epsilon_-(T_-)$.
However, one always uses the same temperature (e.g. the nucleation temperature) to calculate this quantity.
To generalize $\alpha_{\theta}$ without encountering the dependence on the temperature in the broken phase $T_-$,
Ref.~\cite{Giese:2020rtr} expands the thermodynamic quantities around the symmetric phase to derive the corresponding values in the broken phase,
and the constant sound velocity gives the following relation:
\begin{equation}
\frac{v_+}{v_-} = \frac{w_+(v_+v_-/c_-^2 - 1) + \Delta(e - p/c_-^2)}{w_+(v_+v_-/c_-^2 - 1) + v_+v_-\Delta(e - p/c_-^2)} \,\,.
\end{equation}
Note the $\Delta$ represents the differences of various quantities at the temperature $T_+$ henceforth.
Therefore the pseudotrace definition of the strength parameter is given as \cite{Giese:2020rtr}
\begin{equation}
\alpha_{\bar{\theta}} = \frac{\Delta\bar{\theta}}{3w_+},\quad \bar{\theta} = e - p/c_-^2,
\end{equation}
and this can be combined with the matching condition to give
\begin{equation}
v_- = \frac{1}{2v_+(\nu_- - 1)}\left[a \pm \sqrt{a^2 + 4v_+^2(1 - \nu_-)}\right]\,\,, \notag
\end{equation}
\begin{equation}
a = 1 - 3\alpha_{\bar{\theta}} + v_+^2(\nu_- + 3\alpha_{\bar{\theta}} -1)\,\,,\label{vm}
\end{equation}
and 
\begin{equation}
v_+ = \frac{1 - v_-^2(1 - \nu_-) \pm \sqrt{(1 - v_-^2(1 - \nu_-))^2 - 4v_-^2(1 - 3\alpha_{\bar{\theta}})(\nu_- -1 + 3\alpha_{\bar{\theta}})}}{2v_-(\nu_--1+3\alpha_{\bar{\theta}})}\,\,.\label{vp}
\end{equation}

\begin{figure}[t]
	\centering
	
	\subfigure{
		\begin{minipage}[t]{0.5\linewidth}
			\centering
			\includegraphics[scale=0.5]{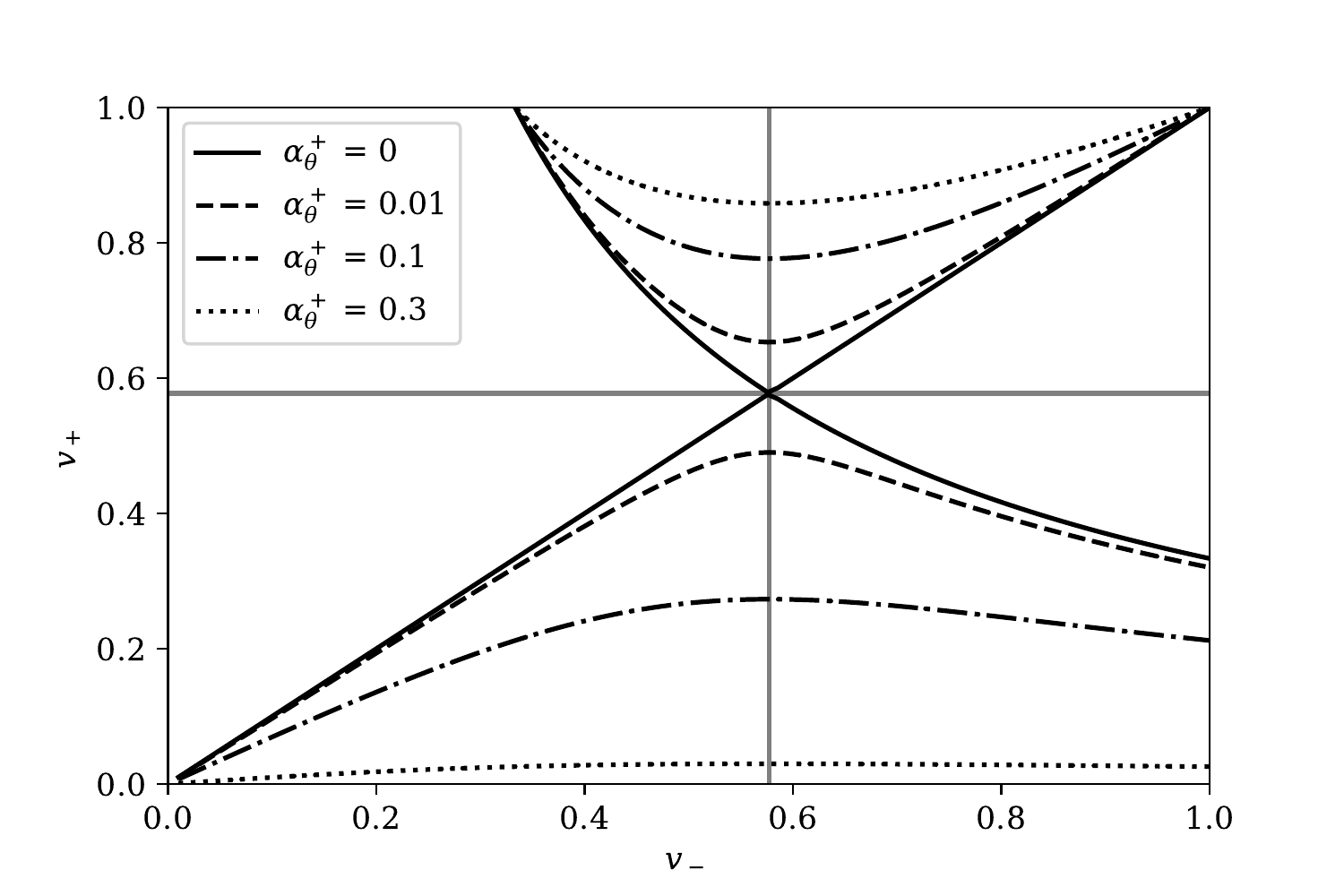}
		\end{minipage}%
	}%
	\subfigure{
		\begin{minipage}[t]{0.5\linewidth}
			\centering
			\includegraphics[scale=0.5]{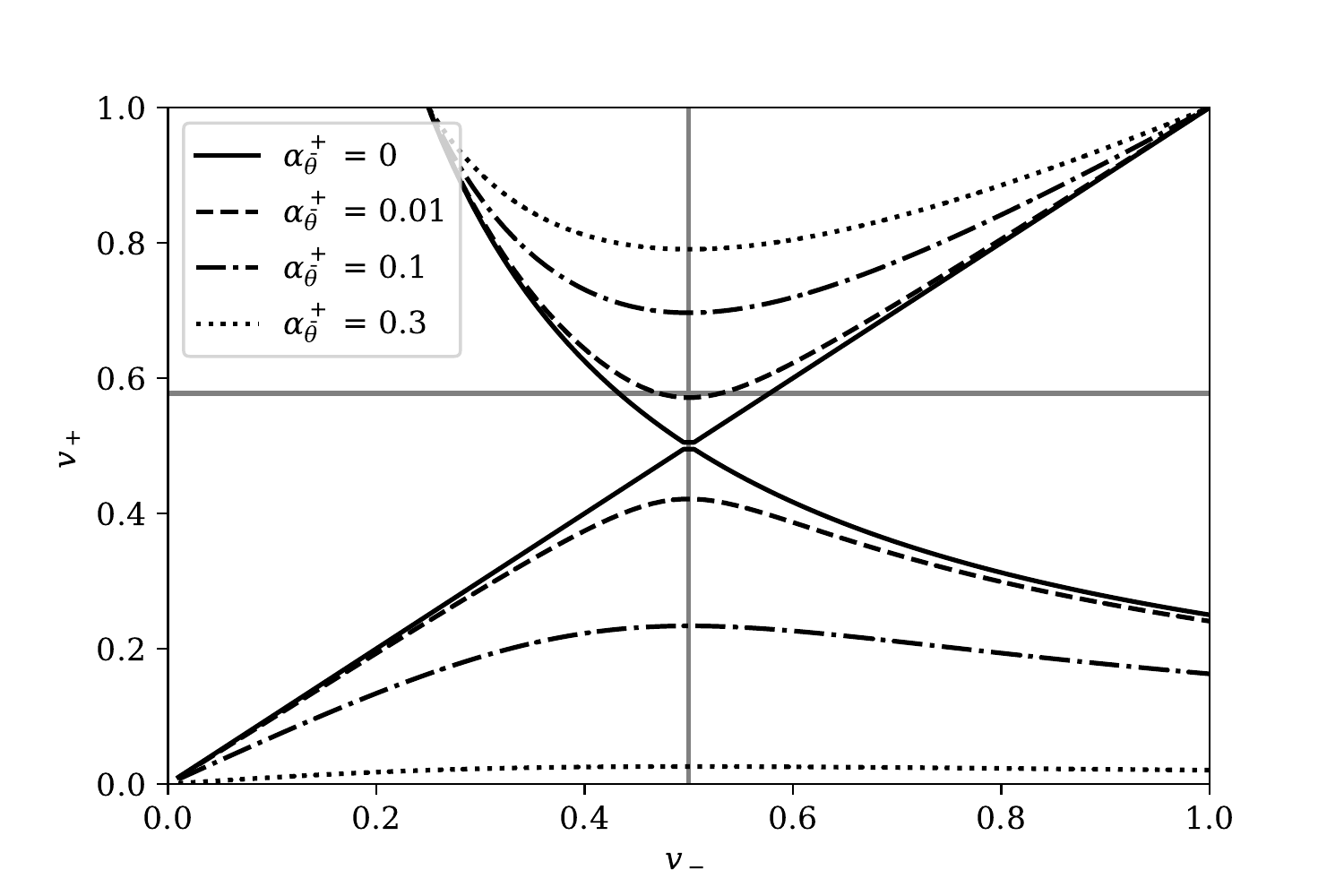}
		\end{minipage}%
	}%
	\centering
	\caption{The fluid velocities $v_+$ and $v_-$ in the reference frame of bubble wall for different definitions and values of phase transition strength parameter. The horizontal and vertical gray lines indicate the sound velocities of symmetric and broken phase. Left panel: the bag model. Right panel: the DSVM with $c_+^2 = 1/3$ and $c_-^2 = 0.25$.}\label{vpvm}
\end{figure}

Hence the $\pm$ signs in Eq.~\eqref{vp} indicate two branches of solutions, and the two branches of solutions indicate two kinds of hydrodynamical processes may occur during the phase transition process.
In the reference frame of the bubble wall, a detonation mode, which is the upper right part of both panels in Fig.~\ref{vpvm}, means the incoming flow is faster than the outgoing flow, $v_- < v_+$,
and a deflagration mode, which is the bottom left part of both panels in Fig.~\ref{vpvm}, indicates the outgoing flow is faster than the ingoing flow, $v_+ < v_-$.
Figure~\ref{vpvm} shows that $v_+$  has a minimum at $v_- = c_-$ in the upper right part of both panels, and maximum at $v_- = c_-$ in the bottom left part of both panels,
and a mode with $v_- = c_-$ is called Jouguet detonation or deflagration.
The corresponding Jouguet  velocity is
\begin{equation}
v_J = \frac{1\pm \sqrt{ 3\alpha_{\bar{\theta}}(1 - c_-^2 + 3c_-^2\alpha_{\bar{\theta}})}}{1/c_- + 3c_-\alpha_{\bar{\theta}}} \,\,,
\end{equation}
where $+$ indicates a Jouguet detonation velocity $v_J^{\rm det}$ and $-$ denotes the Jouguet deflagration velocity $v_J^{\rm def}$.

A further analysis needs to deal with the hydrodynamics of the fluid, which is described by the fluid equation.
Based on the energy-momentum conservation  $\partial_\mu T^{\mu\nu} = 0$, we can also derive the fluid equation as \cite{KurkiSuonio:1984ba,Leitao:2010yw}
\begin{equation}
\partial_t\left[(e + pv^2)\gamma^2\right] + \partial_r\left[(e + p)\gamma^2v\right] = - \frac{j}{r}\left[(e+p)\gamma^2v\right]\,\,,\notag
\end{equation}
\begin{equation}
\partial_t\left[(e+p)\gamma^2v\right] + \partial_r\left[(ev^2 + p)\gamma^2\right] = -\frac{j}{r}\left[(e+p)\gamma^2 v^2 \right]\,\,,
\end{equation}
where $r$ denotes the distance from the symmetry plane, axis, or point, and $t$ is the duration since nucleation, 
and $j = 0, 1, 2$ for the planar, cylindrical, or spherical bubble configuration, respectively.
Since there is no characteristic distance scale in this problem, the solution is similarity solution which depends only on $\xi=r/t$.
Thus, we can derive 
\begin{equation}\label{pp}
(\xi - v)\frac{\partial_{\xi}e}{w} = j\frac{v}{\xi} + \gamma^2(1 - v\xi)\partial_{\xi}v\,\,,\notag
\end{equation}
\begin{equation}
(1 - v\xi)\frac{\partial_{\xi}p}{w} = \gamma^2(\xi - v)\partial_{\xi}v\,\,.
\end{equation}
We assume a spherically symmetric configuration in this work, 
\begin{equation}
(\xi - v)\frac{\partial_{\xi}e}{w} = 2\frac{v}{\xi} + \gamma^2(1 - v\xi)\partial_{\xi}v\,\,,\notag
\end{equation}
\begin{equation}\label{eom}
(1 - v\xi)\frac{\partial_{\xi}p}{w} = \gamma^2(\xi - v)\partial_{\xi}v\,\,. 
\end{equation}
Therefore, the equation that can describe the velocity profile is
\begin{equation}
2\frac{v}{\xi} = \gamma^2(1 - v\xi)\left[\frac{\mu^2}{c_s^2} - 1\right]\partial_{\xi}v\,\,,\label{vprofile}
\end{equation}
where
\begin{equation}
\mu(\xi,v) = \frac{\xi - v}{1 - \xi v}\,\,.
\end{equation}
From Eq.~\eqref{eom} we can also derive the equation for the enthalpy profile,
\begin{equation}
\frac{\partial_{\xi}w}{w} = \left(1 + \frac{1}{c_s^2}\right)\mu\gamma^2\partial_{\xi}v\,\,,
\end{equation}
and the equation for temperature profile
\begin{equation}
\frac{\partial_{\xi}T}{T} = \gamma^2\mu\partial_{\xi}v\,\,.
\end{equation}
Then we can obtain the enthalpy profile 
\begin{equation}
w(\xi) = w_0\exp\left[\int_{v_0}^{v(\xi)}\left(1 + \frac{1}{c_s^2}\right)\gamma^2\mu dv\right]\,\,,\label{wprofile}
\end{equation}
and the temperature profile
\begin{equation}
T(\xi) = T_0\exp\left[\int_{v_0}^{v(\xi)}\gamma^2\mu dv\right]\,\,.\label{Tprof}
\end{equation}
By solving the above fluid equations and obtaining the fluid profiles, we can classify hydrodynamical processes of FOPT into three stable modes: weak detonation, weak deflagration and hybrid \cite{Espinosa:2010hh,Leitao:2010yw,KurkiSuonio:1995pp} .
When the bubble wall velocity ($v_w$) is smaller than the sound velocity of the broken phase ($c_-$), namely $v_w < c_-$, it is called deflagration mode, which forms a shock in front of the bubble walls.
This mode is favored by the electroweak baryogenesis to guarantee sufficient diffusion time.
If the bubble wall velocity is larger than sound velocity of broken phase and smaller than the so-called Jouguet detonation velocity, $c_- < v_w < v_J^{\rm det}(\alpha_{\bar{\theta} n})$, it is called the hybrid mode.
If the bubble wall velocity is even larger than the Jouguet detonation velocity, it is called the detonation mode, which produces a rarefaction wave behind the bubble walls. 
Detonation mode usually produces stronger GW signal. 
The detailed analysis of fluid profile of various hydrodynamical modes are presented in Appendix~\ref{profile}.

\section{Kinetic energy fraction}\label{kef}
After solving the hydrodynamical equations and hence obtaining the fluid profile for a given EOS,
we can further predict the kinetic energy fraction.
The kinetic energy fraction $K$ can be defined as a fraction of the total energy $e_+$ that is contained in the bubble~\cite{Caprini:2019egz},
\begin{equation}
K  \equiv \frac{\rho_{\rm fl}}{e_+}= \frac{3}{e_+ v_w^3} \int \, w(\xi) \, v^2 \gamma^2 \xi^2 d\xi \,\,, \quad \rho_{\rm fl} = \frac{3}{v_w^3}\int\xi^2v^2\gamma^2wd\xi
\end{equation}
In general, the energy spectra of GWs from sound wave mechanism scales as
$h^2\Omega_{\rm GW} \propto K^2$ or $h^2 \Omega_{\rm GW} \propto K^{3/2}$.
In most cases, the kinetic energy fraction $K$ for single bubble could be a good approximation for the whole phase transition process. 
Since from the detailed discussion in the previous section, the enthalpy profile $w(\xi)$ and the fluid velocity profile $v(\xi)$ depend on both the phase transition strength $\alpha$ and the bubble wall velocity $v_w$, 
the single bubble kinetic energy fraction also depends on these parameters.
According to Ref.~\cite{Giese:2020rtr}, we choose the pseudotrace definition of the strength parameter, hence the efficiency parameter can be expressed as
\begin{equation}
\kappa_{\bar{\theta}} = \frac{4\rho_{\rm fl}}{\Delta\bar{\theta}} \,\,,
\end{equation}
and the $K$ can be approximated as 
\begin{equation}
K \approx \left(\frac{\Delta\bar{\theta}}{4e_+}\kappa_{\bar{\theta}}\right)\,\,.\label{fraction}
\end{equation}
To obtain the energy fraction, we need perform a model dependent calculation of the prefactor.

\section{The representative effective model}\label{md}
In this section, we take a representative effective model, the Higgs sextic model~\cite{Zhang:1992fs,Grojean:2004xa,Huang:2015izx,Huang:2016odd,Cao:2017oez} with
the tree-level potential $
V(\phi)=\frac{\mu^2}{2}\phi^2 + \frac{\lambda}{4}\phi^4 + \frac{\kappa}{8\Lambda^2}\phi^6$, 
to make precise predictions on the sound velocity, the 
phase transition dynamics and parameters. 
This effective model is a generic prediction for many new physics models motivated by dark matter, baryogenesis and so on.
From the perspective of standard model  effective field theory, this Higgs sextic operator could naturally appear after the heavy degrees of
freedom are integrated out in the new physics model, such as scalar extended Higgs model or composite Higgs model~\cite{Cao:2017oez}.
This Higgs sextic effective model is still favored by the collider data and electroweak precise measurements after considering other dimension-six operator induced simultaneously 
with the Higgs sextic term.
Basically, the free energy (or effective potential) of a given model can be divided into the following two parts
\begin{equation}\label{freeen}
\mathcal{F}(\phi, T) = V_{\rm eff}(\phi, T) = V_{T=0}(\phi) + V_T(\phi,T)\,\,,
\end{equation}
where $V_{T = 0}(\phi)$ is zero-temperature effective potential, which contains the tree-level potential and the Coleman-Weinberg potential, the one-loop thermal correction in the effective potential is 
\begin{equation}
V_T(\phi,T) = \pm T\int \frac{d^3 k}{(2\pi)^3} \ln(1 \mp e^{-\omega_k/T})=\pm \frac{T^4}{2\pi^2} J_{b/f}\left(\frac{m}{T}\right)\,\,,
\end{equation}
where $\omega_k^2 = m^2 + k^2 $ and the thermal function is
\begin{equation}
J_{b/f}(y^2) = \int_{0}^{\infty}dxx^2\ln\left(1 \mp e^{-\sqrt{x^2+ y^2}}\right)\,\,.
\end{equation}
For bosons, at high temperature limit, $m/T\ll1$, one can expand it as 
\begin{equation}\label{jb}
\begin{split}
J_b(y^2) \approx &-\frac{\pi^4}{45} + \frac{\pi^2}{12}y^2 - \frac{\pi}{6}(y^2)^{3/2} - \frac{1}{32}y^4\ln\frac{y^2}{a_b}\,\,,\\
a_b = &16\pi^2\exp\left(\frac{3}{2} - 2\gamma_E\right)\,\,.
\end{split}
\end{equation}
For fermions, the high temperature expansion is
\begin{equation}\label{jf}
\begin{split}
J_f(y^2) \approx &\frac{7\pi^4}{360} - \frac{\pi^2}{24}y^2 - \frac{1}{32}y^4\ln\frac{y^2}{a_f}\,\,,\\
a_f =& \pi^2\exp\left(\frac{3}{2} - 2\gamma_E\right)\,\,,
\end{split}
\end{equation}
where $y^2 = m^2/T^2$, $\ln a_b = 5.4076$, $\ln a_f = 2.6351$ and $\zeta$ is Riemann $\zeta$-function.
We usually omit the field-independent term in the calculation of phase transition dynamics.
However, in this work we should include the field-independent term $- \frac{a_{\pm}}{3}T^4$, which is the contribution of the zero-order term of the high-temperature expansion.
This is the most important term for the calculation of the sound velocity.
For simplicity, we only study the Higgs sextic term with the leading-order thermal corrections
\begin{equation}
\mathcal{F}(\phi,T)=V_{\rm eff}(\phi,T)\approx - \frac{a_{\pm}}{3}T^4 + \frac{\mu^2 + cT^2}{2}\phi^2 + \frac{\lambda}{4}\phi^4 + \frac{\kappa}{8\Lambda^2}\phi^6  \,\,,
\end{equation}
where $\Lambda/\sqrt{\kappa}$ is the effective cutoff scale and $c$ is the thermal correction
\begin{equation}\label{cdim6}
c=\frac{1}{16}(g^{\prime 2}+3 g^2+4 y_t^2+4 \frac{m_h^2}{v^2}-12 \frac{\kappa v^2}{\Lambda^2}) \,\,.
\end{equation}
 $g^{\prime}$, $g$, $y_t$,  $v$ is the $U(1)$ gauge coupling, $SU(2)$ gauge coupling, top quark Yukawa coupling, and electroweak vacuum expectation value (VEV), respectively. 

We can obtain the sound velocity of broken and symmetric phase, then match this to the DSVM.
In general, the sound velocity is given as
\begin{equation}
c_s^2 = \frac{\partial p/\partial T}{\partial e/\partial T}\,\,,
\end{equation}
where $p = - V_{\rm eff}$ and $e = T\frac{\partial p}{\partial T} -p$.
For this model, we can derive the sound velocity 
\begin{equation}
c_s^2 = \frac{4a_{\pm}T^3 - 3cT\phi^2}{12a_{\pm}T^3 - 3cT\phi^2}\,\,,
\end{equation}
where the value of $\phi$ is zero in symmetric phase.
However, the value for broken phase is 
\begin{equation}
\phi_{\rm true} = \sqrt{\frac{-2\lambda\Lambda^2 + 2\Lambda\sqrt{\lambda^2\Lambda^2 - 3\kappa(\mu^2 + cT^2)}}{3\kappa}}\,\,.
\end{equation}
Then the sound velocity of symmetric phase is
\begin{equation}
c_+^2 = \frac{4a_+T^3}{12a_+^2T^3} = \frac{1}{3}\,\,.
\end{equation}
For the broken phase, the corresponding sound velocity is
\begin{equation}
c_-^2 = \frac{4a_-T^3 - 3cT\phi_{\rm true}^2}{12a_-T^3 - 3cT\phi_{\rm true}^2}\,\,,
\end{equation}
and the phase transition strength parameter can be obtained as 
\begin{equation}
\alpha_{\bar{\theta} n} = \frac{(1 + 1/c_-^2)\Delta V_{\rm eff} - T\frac{\partial\Delta V_{\rm eff}}{\partial T}}{3(1 + c_+^2)\rho_{\rm R}},\quad \rho_{\rm R} = a_+T_+^4\,\,.
\end{equation}
From the above equations, we find the sound velocity is actually temperature-dependent.
Hence choosing a different temperature can give a different efficiency parameter.

The conventional definition of the  phase transition strength parameter based on the bag model of  EOS can be written as
\begin{equation}
\alpha_{\theta} = \frac{\Delta V_{\rm eff} - \frac{T}{4}\frac{\partial\Delta V_{\rm eff}}{\partial T}}{\rho_{\rm R}} \,\,.
\end{equation}
This definition is based on the same sound velocity (i.e. $c_{\pm} = 1/\sqrt{3}$) approximation in the broken and symmetric phase.
For the bag model and the corresponding definition of phase transition strength parameter,  we can derive the kinetic energy fraction as 
\begin{equation}
K_{\theta} = \frac{\alpha_{\theta}\kappa_{\theta}}{1 + \alpha_{\theta}}\,\,.
\end{equation}
Considering different sound velocities in both phases, the more realistic kinetic energy fraction should be modified as Eq.~\eqref{fraction}.
For the Higgs sextic  model, $e_+ = a_+T^4$ in symmetric phase, we can obtain
\begin{equation}
K_{\bar{\theta}} = \frac{3}{4}(1 + c_+^2)\alpha_{\bar{\theta}}\kappa_{\bar{\theta}}\,\,,
\end{equation}
where $c_+^2 = 1/3$.
Hereafter, we use $\theta$ to represent the parameters calculated using the bag model with $c_{\pm}^2=1/3$,
and we use $\bar{\theta}$ to represent the quantities derived from the DSVM of EOS with different sound velocities in the symmetric and broken phase.

In Table~\ref{tpara}, we list the important phase transition parameters of different benchmark points.
We only consider the detonation case for simplicity with  the bubble wall velocity $v_w = 0.95$.
From this table, we can observe that the sound velocity of the broken phase is different for different cutoff scales. 
Basically, lower cutoff gives smaller sound velocity, and the deviation from the bag model is more obvious.
The precise definition of these phase transition parameters
can be found in our previous study~\cite{Wang:2020jrd},  and we give a brief explanation of these parameters in the following.

$T_n$ is the nucleation temperature at which one bubble is nucleated in one Hubble radius.
$34\%$ of false vacuum has been converted to true vacuum at  the percolation temperature $T_p$, at which a large numbers of bubbles have collided and percolated.
$\alpha_{\bar{\theta} n}$ is the phase transition strength given by the DSVM of EOS at the nucleation temperature.
$\alpha_{\theta n}$ is the phase transition strength given by the bag model at the nucleation temperature.
At the percolation temperature, $\alpha_{\bar{\theta} p}$ and $\alpha_{\theta p}$ are the phase transition strength derived by the DSVM and the bag model respectively.
$\tilde{\beta}_n$ and $\tilde{\beta}_p$ represent the values of phase transition duration that are obtained at the nucleation and percolation temperature respectively.
$HR_p$ quantifies the mean bubble separation at the percolation temperature.
We can basically use the time duration parameter $\tilde{\beta}$ to approximate it \cite{Wang:2020jrd}.
In the broken phase, the values of sound velocity square for nucleation and percolation temperature are denoted by  $c_{-n}^2$ and $c_{-p}^2$.
According to the DSVM, we can derive the efficiency parameter at nucleation and percolation temperature as $\kappa_{\bar{\theta}n}$ and $\kappa_{\bar{\theta}p}$.
Then we can subsequently obtain the corresponding kinetic energy fraction, $K_{\bar{\theta}n}$ and $K_{\bar{\theta}p}$.
For the bag model, we can also derive the efficiency parameter,
$\kappa_{\theta n}$ and $\kappa_{\theta p}$, and the kinetic fraction, $K_{\theta n}$ and $K_{\theta p}$, at the nucleation and percolation temperature.

\begin{table}[t]\small%
	\centering
	\begin{tabular}{c|cccccc}
		\hline\hline
		& $BP_1$ & $BP_2$ & $BP_3$ & $BP_4$ & $BP_5$ & $BP_6$ \\
		\hline
		$\Lambda/\sqrt{\kappa}$~[GeV] & 620&610&600&590&587&586\\
		$T_n$~[GeV]&65.286&60.153&53.581&43.454&38.118&35.399\\
		$T_p$~[GeV] &64.032&58.751&51.738&40.538&34.364&30.795\\
		$\alpha_{\bar{\theta} n}$&0.0442&0.0634&0.106&0.269&0.490&0.693\\
		$\alpha_{\theta n}$ &0.0435&0.0617&0.101&0.247&0.430&0.588\\
		$\alpha_{\bar{\theta} p}$&0.0491&0.0717&0.126&0.375&0.808&1.378\\
		$\alpha_{\theta p}$&0.0481&0.0695&0.120&0.336&0.674&1.066\\
		$\tilde{\beta}_n$&1023.948&722.848&438.429&177.375&97.708&62.392\\
		$\tilde{\beta}_p$&799.409&652.238&326.527&129.352&50.517&11.879\\
		$HR_p$&0.00490&0.00884&0.0117&0.0276&0.0628&0.114\\
		$c_{+n}^2$&1/3&1/3&1/3&1/3&1/3&1/3\\
		$c_{+p}^2$&1/3&1/3&1/3&1/3&1/3&1/3\\
		$c_{-n}^2$&0.3155&0.3115&0.3048&0.2875&0.2716&0.2602\\
		$c_{-p}^2$&0.3145&0.3102&0.3024&0.2795&0.2548&0.2317\\
		$\kappa_{\bar{\theta}n}$&0.0578&0.0784&0.118&0.224&0.306&0.352\\
		$\kappa_{\bar{\theta}p}$&0.0633&0.0869&0.135&0.269&0.372&0.424\\
		$K_{\bar{\theta}n}$&0.00255&0.00497&0.0125&0.0603&0.150&0.244\\
		$K_{\bar{\theta}p}$&0.00311&0.00623&0.0170&0.101&0.301&0.584\\
		$\kappa_{\theta n}$&0.0626&0.0865&0.134&0.276&0.403&0.485\\
		$\kappa_{\theta p}$&0.0688&0.0964&0.156&0.344&0.521&0.641\\
		$K_{\theta n}$&0.00261&0.00503&0.0123&0.0548&0.121&0.179\\
		$K_{\theta p}$&0.00316&0.00626&0.0167&0.0864&0.210&0.330\\
		\hline\hline
	\end{tabular}
	\caption{Phase transition parameters of different benchmark points for different definitions and different models of EOS. We only consider the detonation case with the bubble wall velocity  $v_w = 0.95$ for simplicity.}\label{tpara}
\end{table}

\begin{figure}
	\centering
	
	\subfigure{
		\begin{minipage}[t]{1\linewidth}
			\centering
			\includegraphics[scale=0.8]{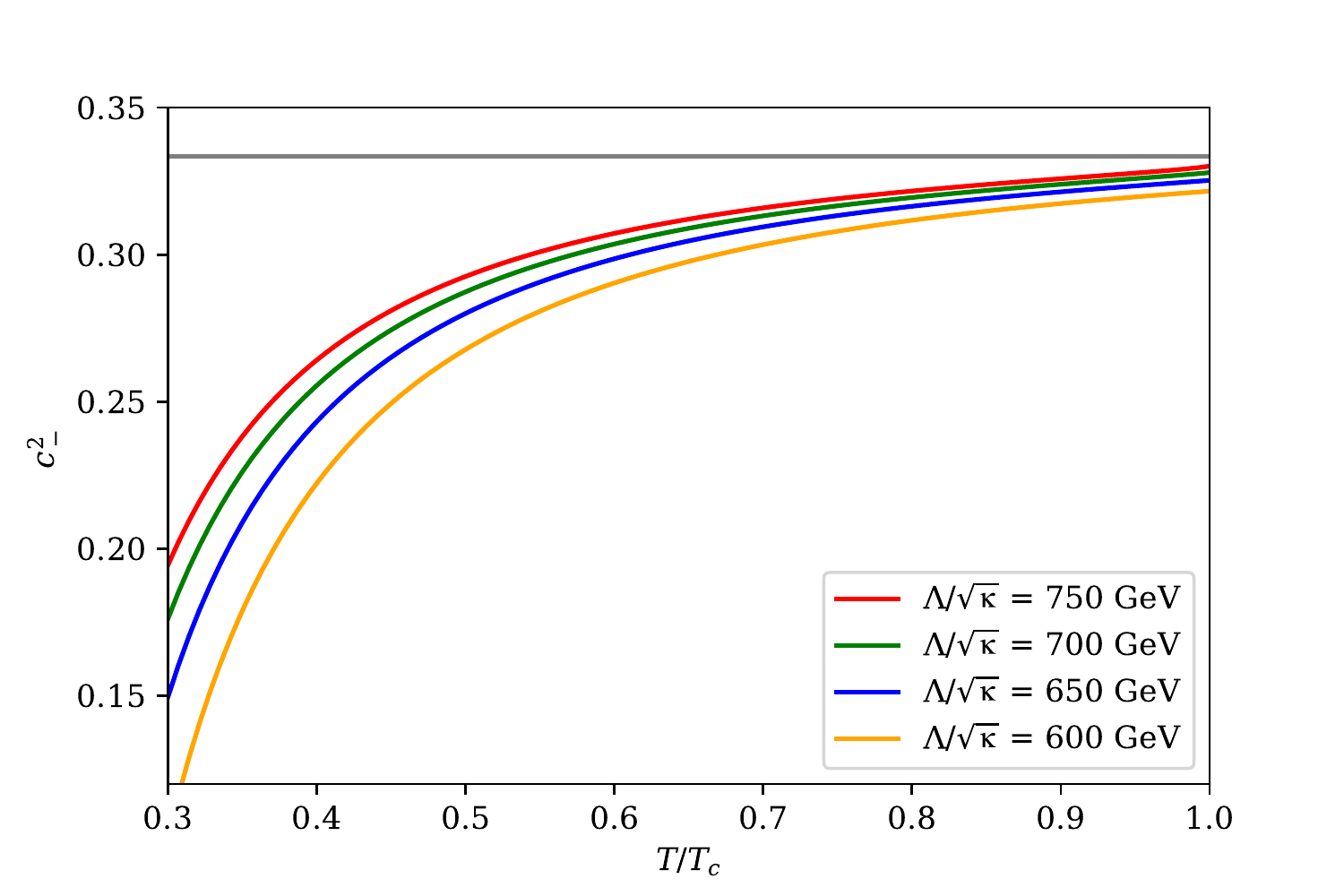}
		\end{minipage}%
	}%
	\centering
	\caption{The evolution of the  sound velocity with the temperature in the broken phase for different cutoff scales. The horizontal gray line represents $c_-^2 = 1/3$ and the sound velocity of the symmetric phase is $c_+^2 = 1/3$.}\label{cmsq}
\end{figure}

In Fig.~\ref{cmsq}, for this Higgs sextic effective model, we show the evolution of the sound velocity with the temperature in the broken phase for different cutoff scales. 
For lower cutoff scales, the supercooling is more significant. Hence the nucleation temperature and percolation temperature become much lower than the masses of $W$
boson, $Z$ boson and Higgs boson for strong supercooling and ultra supercooling cases. These situation would lead to deviation from pure radiation phase. Thus, 
with the decreasing of the temperature, the sound velocity  also decreases, and is smaller than the sound velocity in pure radiation phase (i.e. $c_s=1/\sqrt{3}$).
Since the kinetic energy fraction or EOS depends on the sound velocity in two phases, any deviation from $1/\sqrt{3}$ might have some effects on the kinetic energy fraction.
In our previous study~\cite{Wang:2020jrd}, we classify the FOPT into four classes, slight supercooling for $\alpha<0.1$, mild supercooling for $0.1<\alpha<0.5$, strong supercooling for $0.5<\alpha<1$ and ultra supercooling for $\alpha>1$.  
For strong supercooling and ultra supercooling, the nucleation temperature is obviously smaller than the critical temperature.
Therefore, the sound velocity deviation, which can be observed in Table~\ref{tpara}, becomes more obvious for strong supercooling and ultra supercooling.

\section{Gravitational wave signals and signal-to-noise ratio}\label{gwsnr}
In this section, we precisely calculate the GW spectra for the Higgs sextic effective scenario with the accurate results of the kinetic energy fraction in the bag model and the DSVM.
According to previous studies, there are three known mechanisms to produce GWs during a FOPT, namely,  the bubble collision, turbulence and sound wave.
In most realistic models, sound wave mechanism produces stronger signals than bubble collision and turbulence.
For sound wave, the energy spectrum of GW scales as 
$\Omega_{\rm GW} \propto K^2$ or
$\Omega_{\rm GW} \propto K^{3/2}$ for $H_*  \tau_{sh}<1$,
where $\tau_{sh}$ is the shock formation time \cite{Caprini:2019egz,Wang:2020jrd,Hindmarsh:2017gnf}.

More precisely, for bubble wall velocity away from the Jouguet detonation velocity, we have~\cite{Caprini:2019egz,Hindmarsh:2017gnf}
\begin{equation}\label{swa}
\frac{d \Omega_{GW}}{d \ln f} = 
0.687
F_{\text{GW}} K^2 (H_* R_*/c_s)  C\left(\frac{f}{f_p}\right)\eta \, ,
\end{equation}
where the coefficient $\eta \sim 10^{-2}$, and 
\begin{equation}
F_{\text{GW}} = (3.57 \pm 0.05)\times 10^{-5}  \left(\frac{100}{g_*}\right)^{\frac{1}{3}}.
\label{fgw}
\end{equation}
The spectral shape function is
\begin{equation}
\label{shapegw}
C(s) = s^3\left(\frac{7}{4+3s^2}\right)^\frac{7}{2}\,\,,
\end{equation}
with the peak frequency
\begin{equation}
f_{p}
\simeq 26 \left( \frac{1}{H_*  R_*} \right) \left( \frac{z_p}{10} \right) 
\left(  \frac{T_{*}}{100 \, \text{GeV}} \right) \left(  \frac{g_*}{100}  \right)^{\frac{1}{6}} \; \mu\text{Hz}.
\end{equation}
$T_*$ is the characteristic temperature of the GW production. 
It should be  the percolation temperature $T_p$ and could be approximated as the nucleation temperature $T_n$ if the supercooling is not strong~\cite{Wang:2020jrd}.
$H_*$ and  $R_*$ is the Hubble parameter and the mean bubble separation calculated at the nucleation temperature $T_n$ or the percolation temperature $T_p$.
The quantity $z_p \approx 10$ is determined from simulations~\cite{Caprini:2019egz,Hindmarsh:2017gnf}.
$g_*$ is the effective degree of freedom at $T_*$.
When we consider the sound wave contribution, the contributions from  bubble collision and turbulence are much smaller, and can be neglected.

If $\tau_{sh} H_* <1$, the GW spectra is suppressed and can be written as~\cite{Caprini:2019egz,Hindmarsh:2017gnf}.
\begin{equation}\label{swsp}
\frac{d \Omega_{GW}}{d \ln f} = 
0.687
F_{GW} K^{3/2} (H_* R_*/\sqrt{c_s})^2  C \left(\frac{f}{f_p} \right)\eta.
\end{equation}
For the suppressed sound wave spectra, the contributions from turbulence and bubble collision may not be  negligible. 
Recent study \cite{Guo:2020grp} shows another possible suppression for the sound wave spectra.
Then we might include the GW spectra from turbulence and bubble collisions.

\begin{figure}[t]
	\centering
	
	\subfigure{
		\begin{minipage}[t]{0.5\linewidth}
			\centering
			\includegraphics[scale=0.5]{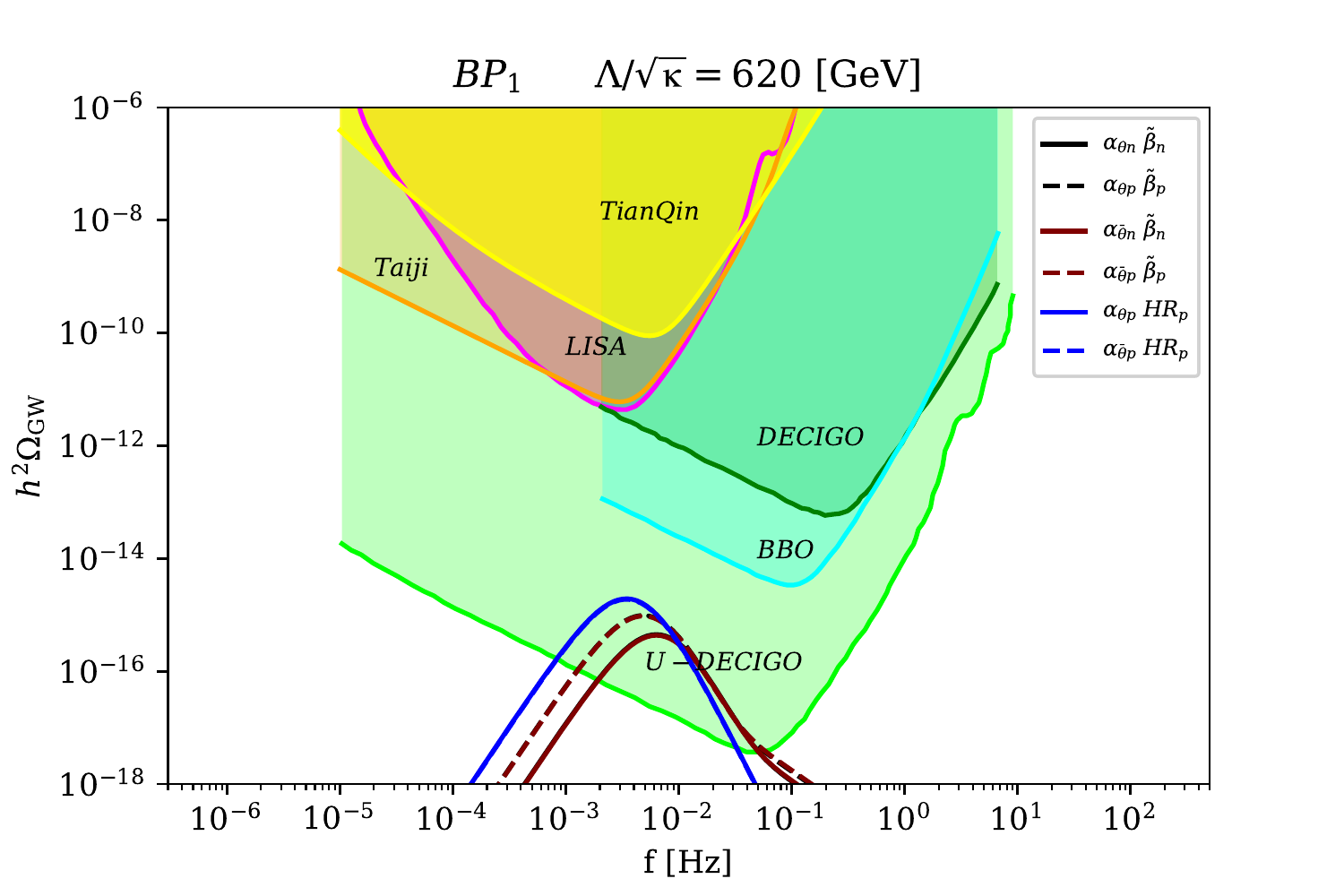}
		\end{minipage}%
	}%
	\subfigure{
		\begin{minipage}[t]{0.5\linewidth}
			\centering
			\includegraphics[scale=0.5]{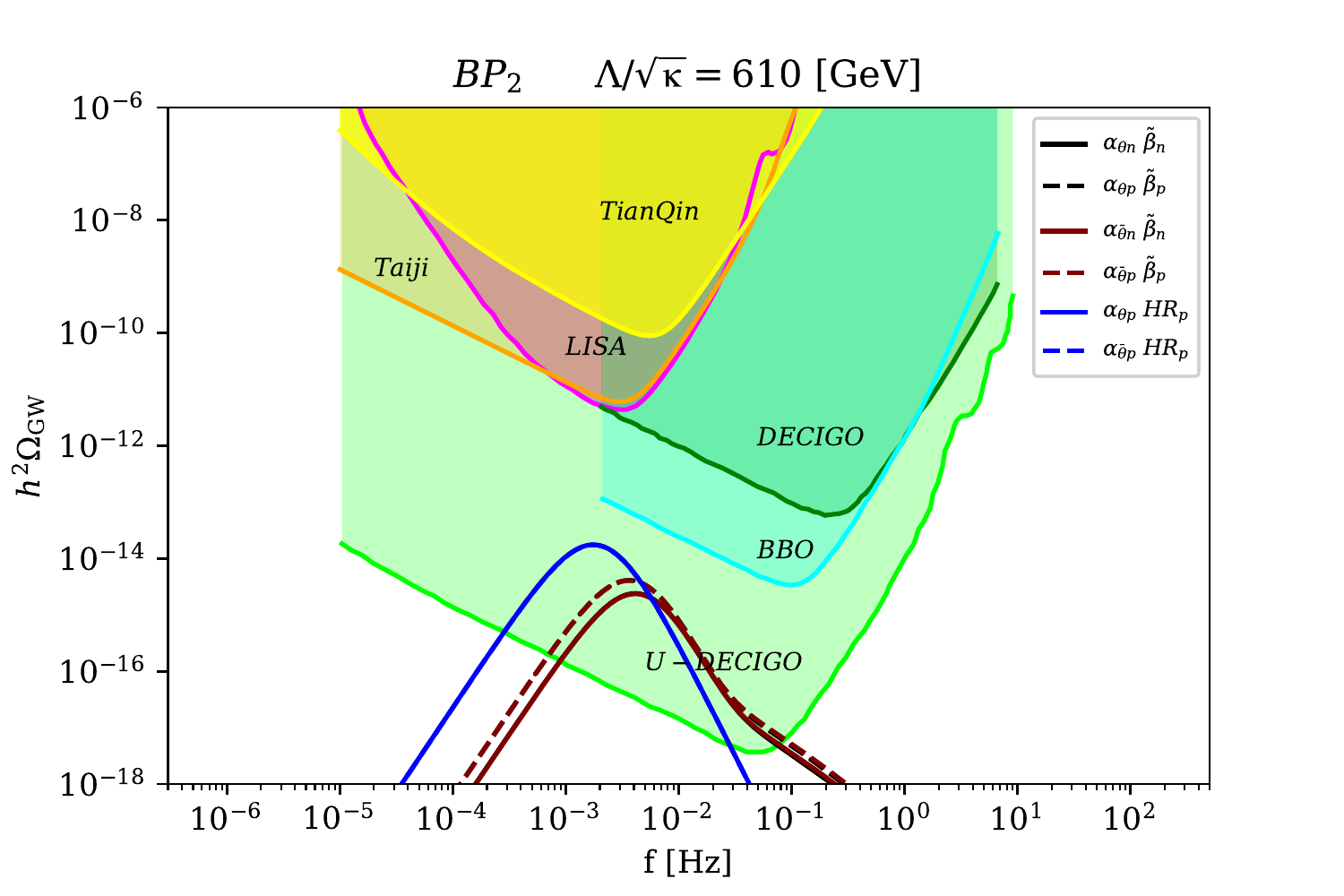}
		\end{minipage}%
	}%
	\quad
	\subfigure{
		\begin{minipage}[t]{0.5\linewidth}
			\centering
			\includegraphics[scale=0.5]{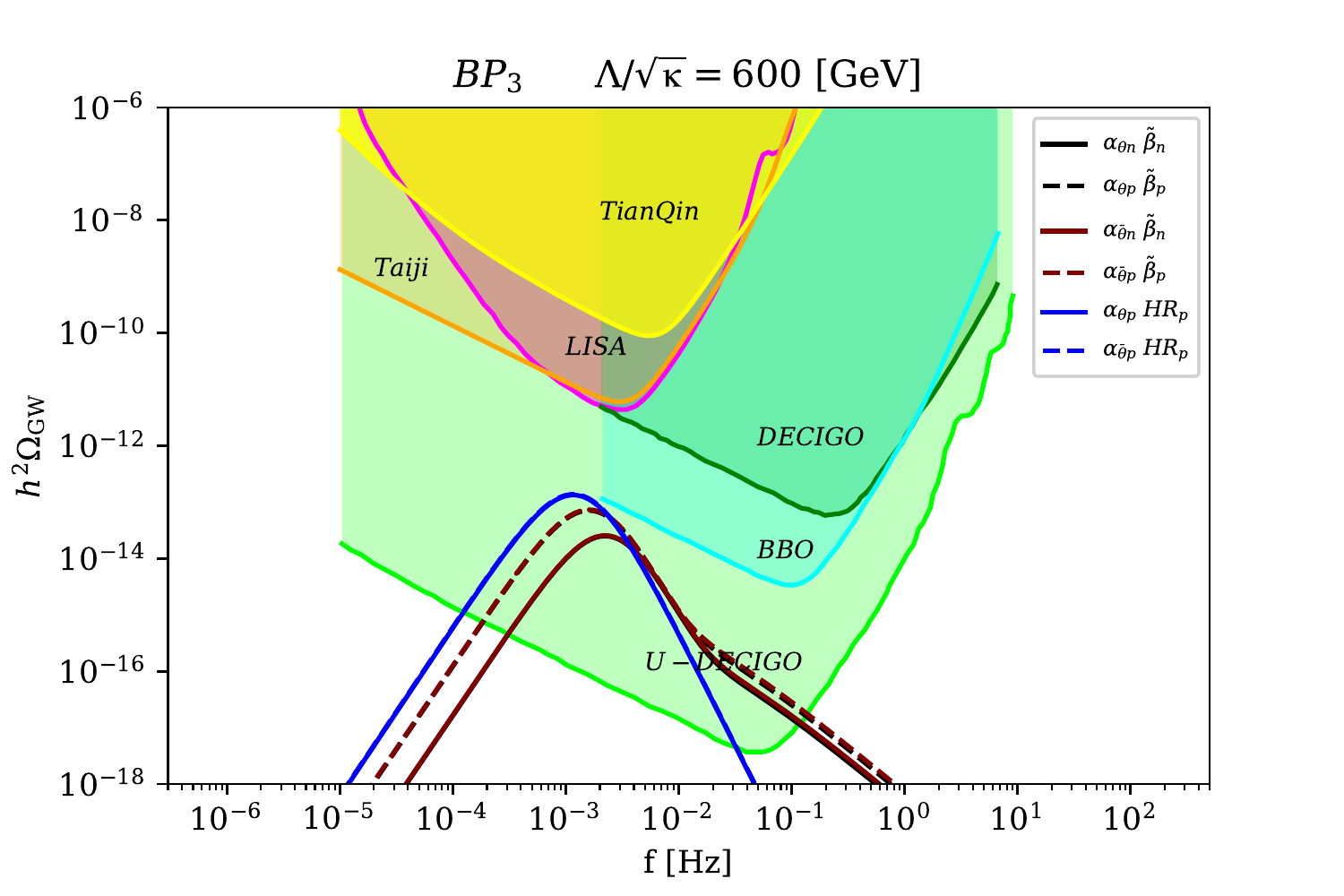}
		\end{minipage}
	}%
	\subfigure{
		\begin{minipage}[t]{0.5\linewidth}
			\centering
			\includegraphics[scale=0.5]{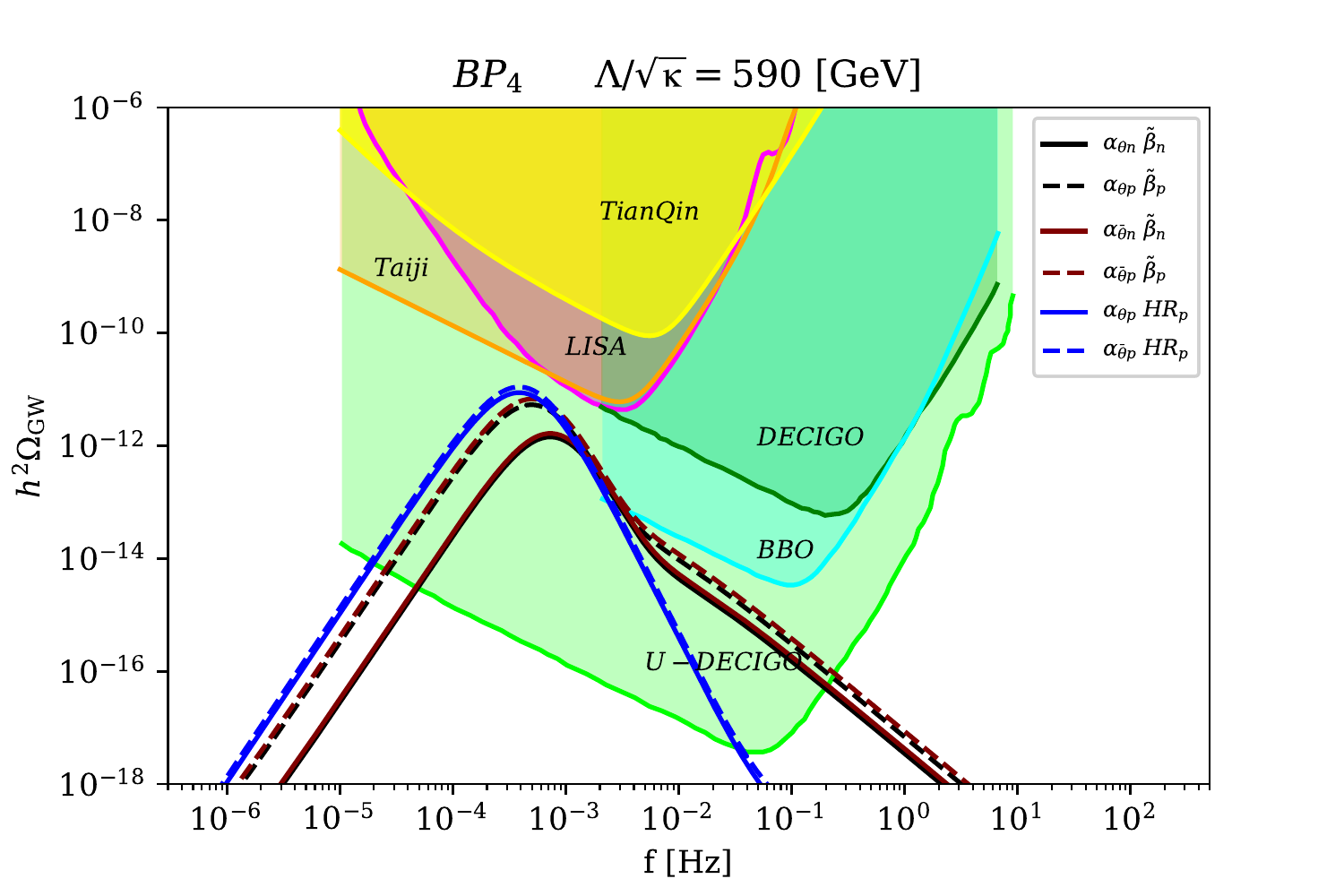}
		\end{minipage}
	}%
	\quad
	\subfigure{
		\begin{minipage}[t]{0.5\linewidth}
			\centering
			\includegraphics[scale=0.5]{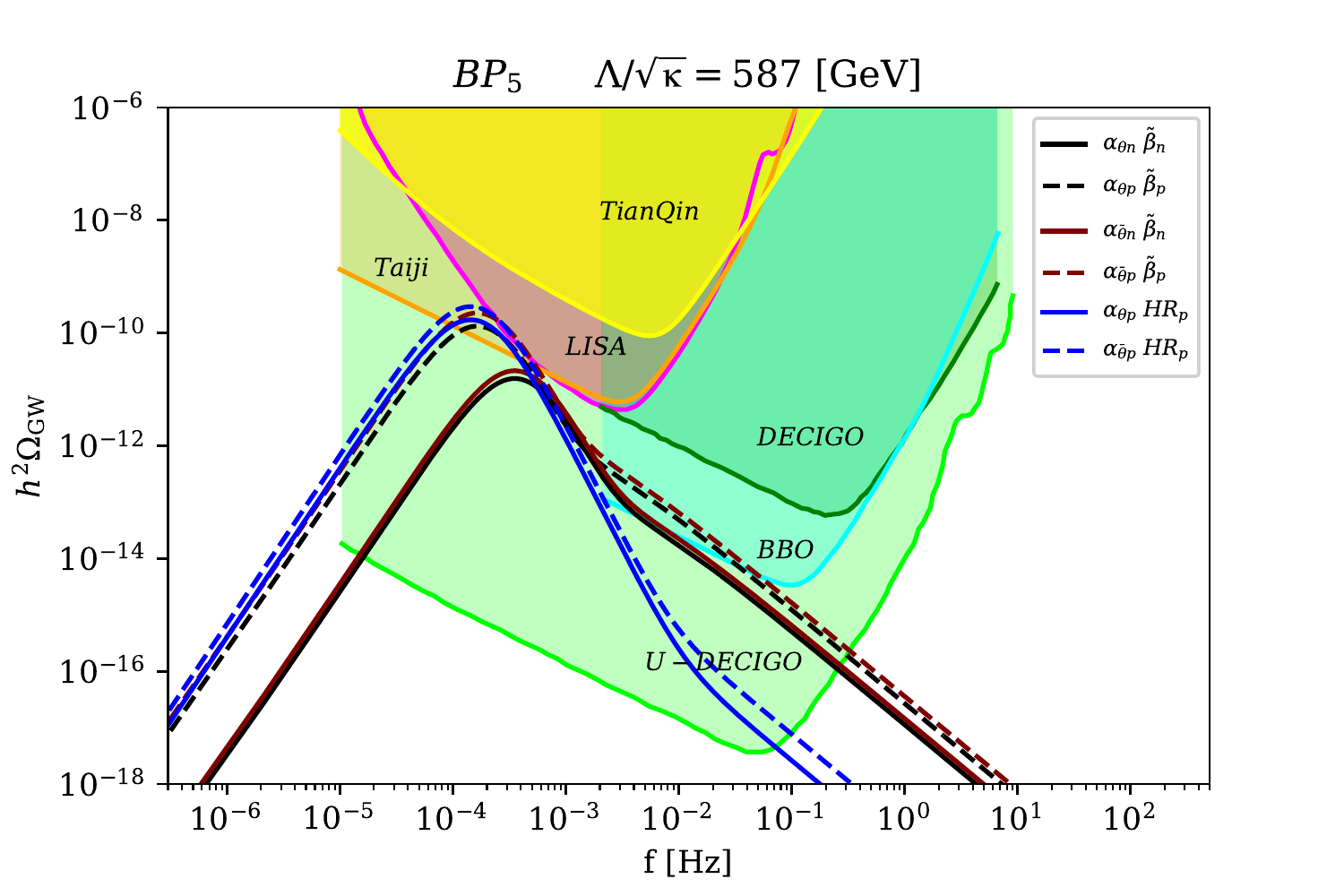}
		\end{minipage}
	}%
	\subfigure{
		\begin{minipage}[t]{0.5\linewidth}
			\centering
			\includegraphics[scale=0.5]{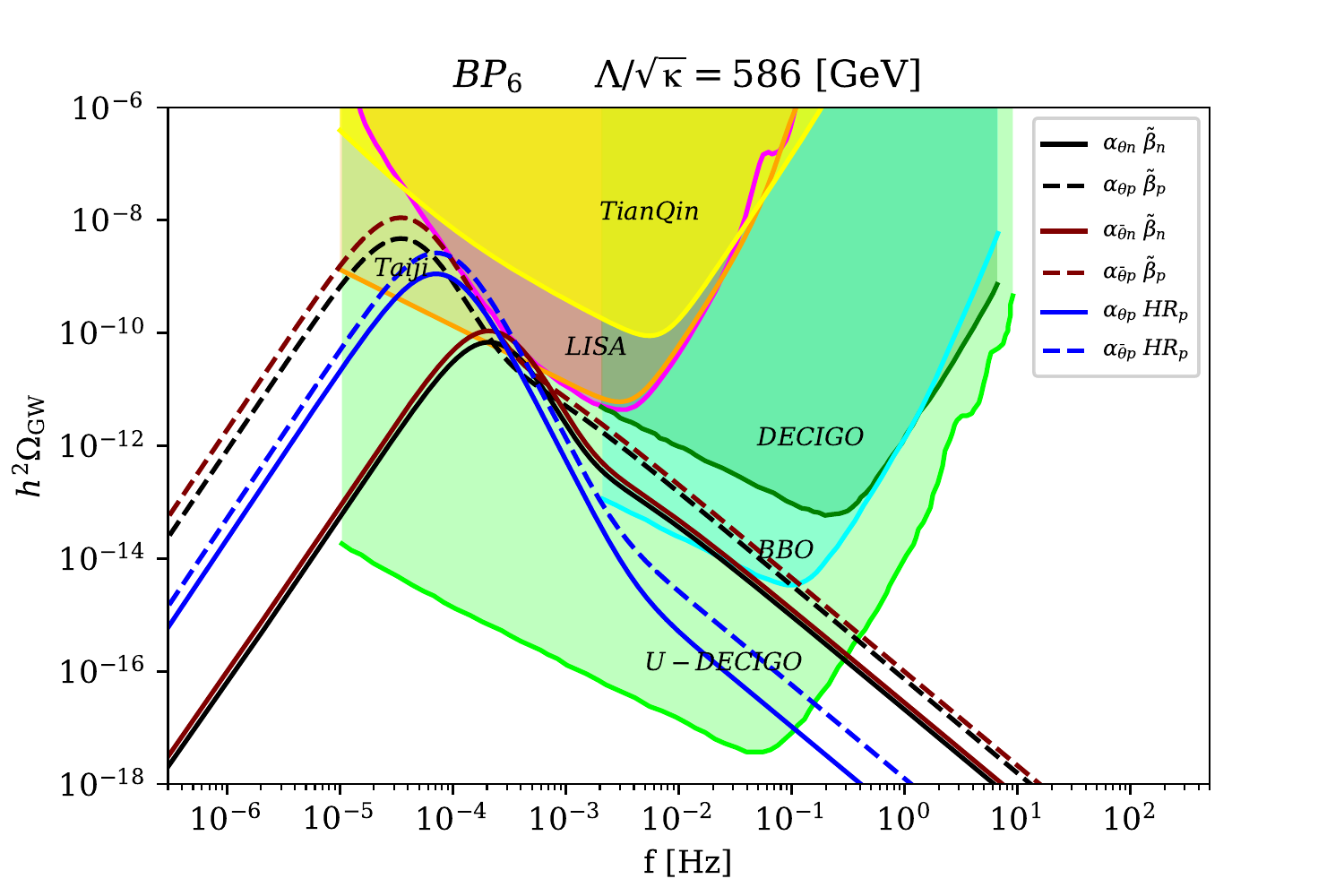}
		\end{minipage}
	}%
	\caption{GW spectra for the six benchmark points with different definitions of phase transition parameters and different models of EOS. The colored regions represent the expected sensitivities for the future GW experiments,
LISA, TianQin, Taiji, DECIGO,
U-DECIGO,  and BBO.}
	\label{gw1}
\end{figure}

According to the numerical calculations, we find $\tau_{sh} H_* <1$ for all benchmark points of the Higgs sextic model. 
Thus, we should use the suppressed GW spectra in Eq.~\eqref{swsp},
and we find the turbulence can have non-negligible contribution.
The GW spectrum from bubble collision is too small and can be neglected in our numerical results.
Based on the kinetic energy fraction and other phase transition parameters derived from the bag model and the DSVM, we show the GW spectra of the six different benchmark points, which are derived by different combinations of the phase transition parameters and models of EOS, in Fig.~\ref{gw1}.
The colored regions represent the expected sensitivity of the future GW experiments,
LISA~\cite{Audley:2017drz,Caprini:2015zlo,Caprini:2019egz,LISA:documents},
TianQin~\cite{Luo:2015ght,Hu:2018yqb,Mei:2020lrl}, Taiji~\cite{Hu:2017mde,Guo:2018npi}, DECIGO~\cite{Seto:2001qf,Kawamura:2011zz},
U-DECIGO~\cite{Kudoh:2005as}, and  BBO~\cite{Corbin:2005ny}, respectively.
The signals for most of the benchmark are within the sensitivities of  LISA, Taiji, DECIGO, U-DECIGO, and BBO.
The GW spectra, derived with different combinations of phase transition parameters at different characteristic temperatures, could have a hierarchy that can reach at most two orders of magnitude. 
However the spectra show that there are slight differences between the same parameter combinations that are derived by different EOS for the same temperature.

To clearly quantify whether the signal is detectable for a given GW experiment, we should also estimate the SNR for each case with the following formula:
\begin{equation}
  \text{SNR} = \sqrt{\mathcal{T} \int_{f_\text{min}}^{f_\text{max}}
    \mathrm{d}f \left[ \frac{h^2 \Omega_\text{GW} (f) }{ h^2
        \Omega_\text{det}(f)} \right]^2 } \, ,
  \label{snr}
\end{equation}
where $\mathcal{T}$ is the total observation time and $h^2 \Omega_\text{det}(f)$ is the nominal sensitivity of a
given GW experiment configuration to cosmological sources.
For simplicity, we assume 4 years  mission duration time  with a duty cycle of 75\%$\mathcal{T}$,  and  take  $\mathcal{T}\simeq 9.46\times 10^7\,$s.
In Table~\ref{tsnr1} and Table~\ref{tsnr2}, we list the SNR of $BP_5$ and $BP_6$ for  different experiment configurations with different  combinations of phase transition parameters, respectively.
We can see that LISA, TianQin, Taiji are capable of detecting the signals for enough observation time.
The SNR of $ BP_6$ is larger than the SNR in $BP_5$.
For each benchmark point, there exists obvious modification to the SNR for different parameter combination. 
Therefore, to obtain more precise predictions on the SNR of the GW signal, it is important to choose a proper phenomenological EOS of the plasma, which can approximately describe the phase transition process, and the parameter combinations at appropriate temperature.

\begin{table}[t]\small%
	\centering
	\begin{tabular}{ccccccc}
		\hline\hline
		&$\alpha_{\theta n}$ $\tilde{\beta}_n$&$\alpha_{\theta p}$ $\tilde{\beta}_p$&$\alpha_{\bar{\theta} n}$ $\tilde{\beta}_n$&$\alpha_{\bar{\theta} p}$ $\tilde{\beta}_p$&$\alpha_{\theta p}$ $HR_p$&$\alpha_{\bar{\theta} p}$ $HR_p$\\
		\hline
		$\rm SNR_{(\rm LISA)}$&7.949&16.930&10.913&28.836&16.009&27.468\\
		$\rm SNR_{(\rm Taiji)}$&14.760&58.607&20.271&100.343&66.216&113.609\\
		$\rm SNR_{(\rm TianQin)}$&0.452&1.506&0.620&2.576&1.629&2.794\\
		\hline\hline
	\end{tabular}
	\caption{The SNR of $BP_5$ for  different experiment configurations with different combinations of phase transition parameters and models of EOS.}\label{tsnr1}
\end{table}

\begin{table}[t]\small%
	\centering
	\begin{tabular}{ccccccc}
		\hline\hline
		&$\alpha_{\theta n}$ $\tilde{\beta}_n$&$\alpha_{\theta p}$ $\tilde{\beta}_p$&$\alpha_{\bar{\theta} n}$ $\tilde{\beta}_n$&$\alpha_{\bar{\theta} p}$ $\tilde{\beta}_p$&$\alpha_{\theta p}$ $HR_p$&$\alpha_{\bar{\theta} p}$ $HR_p$\\
		\hline
		$\rm SNR_{(\rm LISA)}$&14.230&15.368&22.470&26.382&17.367&40.816\\
		$\rm SNR_{(\rm Taiji)}$&38.666&427.813&61.208&1000.501&213.123&500.668\\
		$\rm SNR_{(\rm TianQin)}$&1.060&5.569&1.678&12.934&3.973&9.333\\
		\hline\hline
	\end{tabular}
	\caption{The SNR  of  $BP_6$ for  different experiment configurations with different combinations of phase transition parameters and models of EOS.}\label{tsnr2}
\end{table}

\section{Discussion}\label{dis}
\subsection{The effect of reheating}
\begin{figure}[t]
	\centering
	
	\subfigure{
		\begin{minipage}[t]{0.5\linewidth}
			\centering
			\includegraphics[scale=0.5]{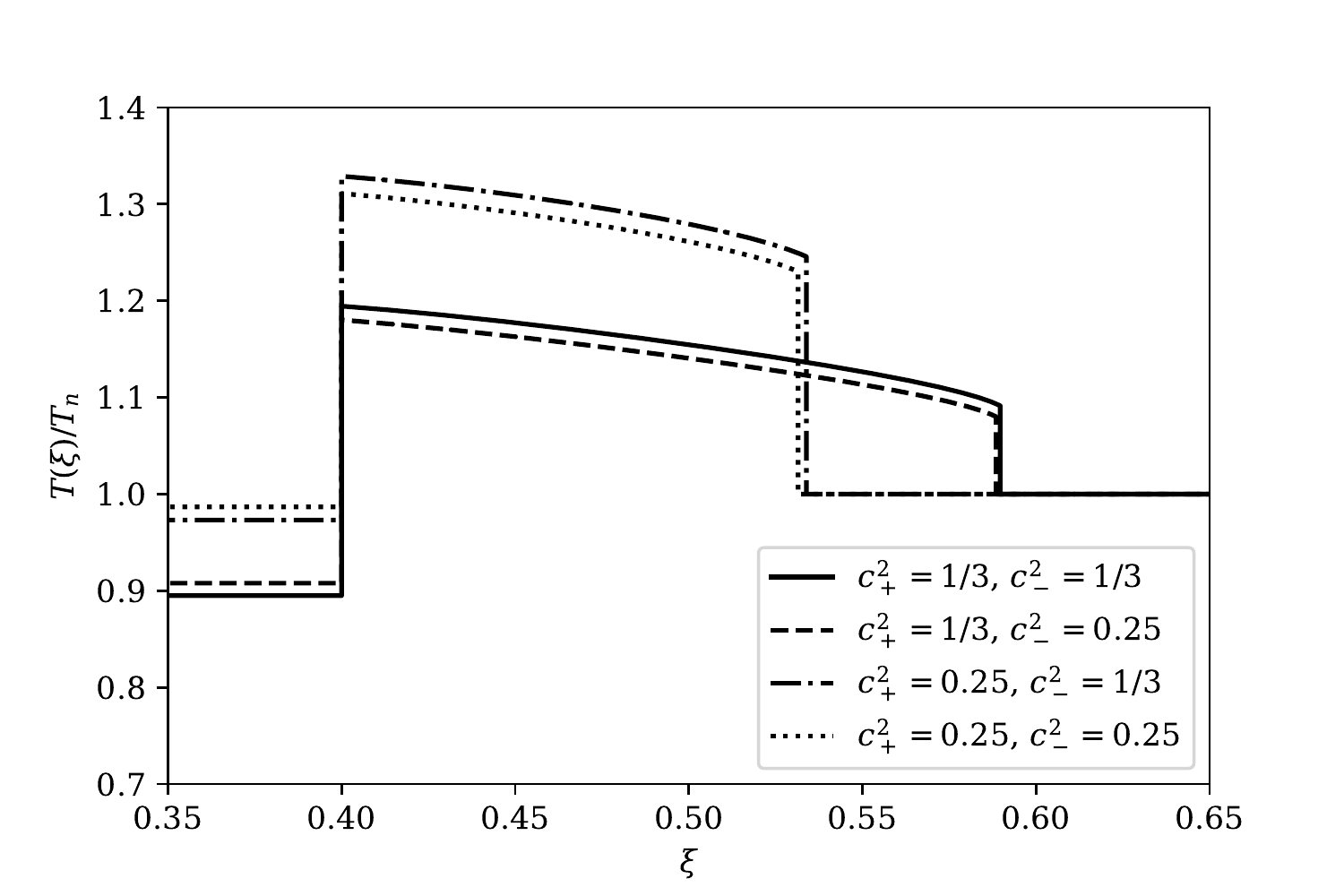}
		\end{minipage}%
	}%
	\subfigure{
		\begin{minipage}[t]{0.5\linewidth}
			\centering
			\includegraphics[scale=0.5]{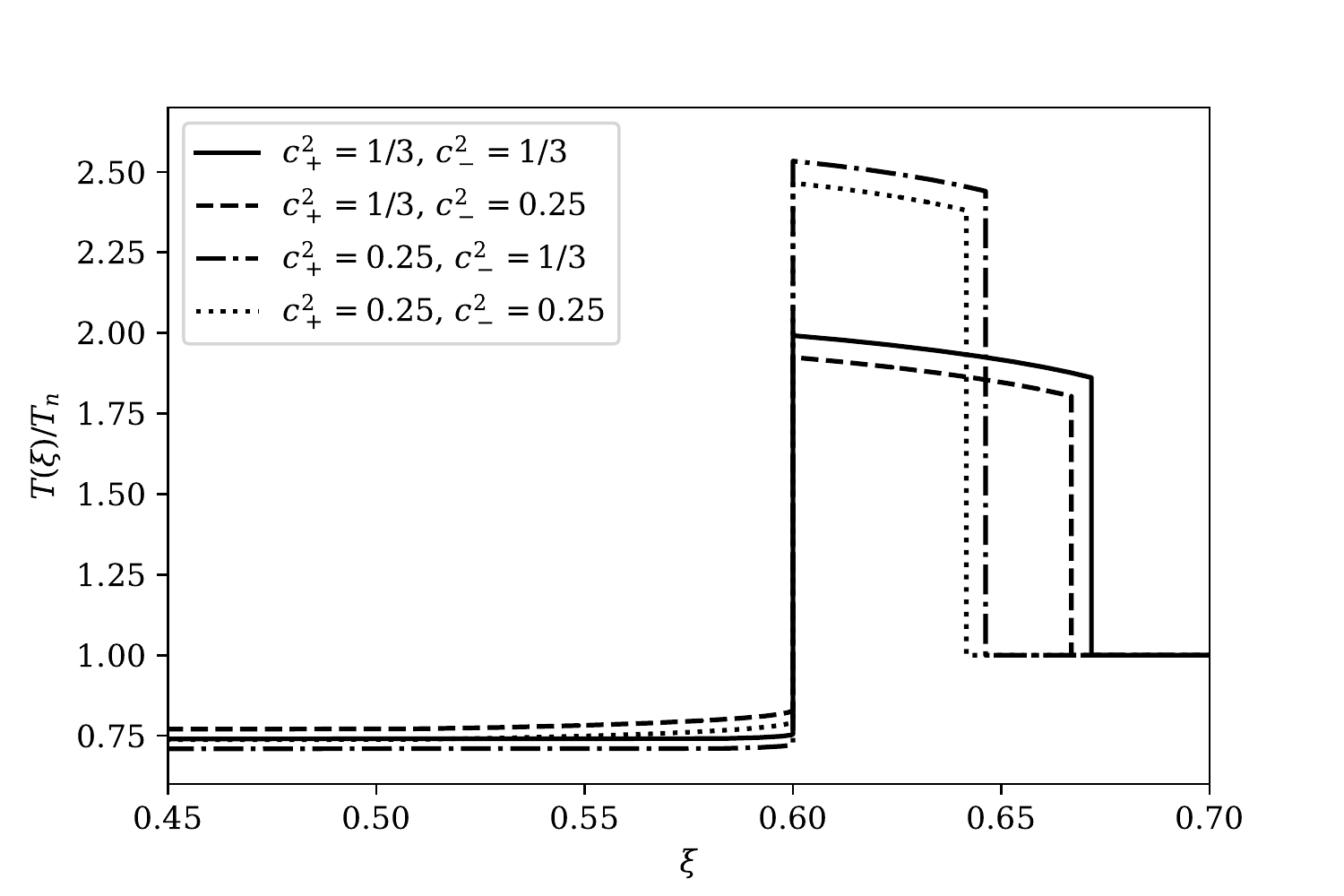}
		\end{minipage}%
	}%
    \quad
	\subfigure{
		\begin{minipage}[t]{0.5\linewidth}
			\centering
			\includegraphics[scale=0.5]{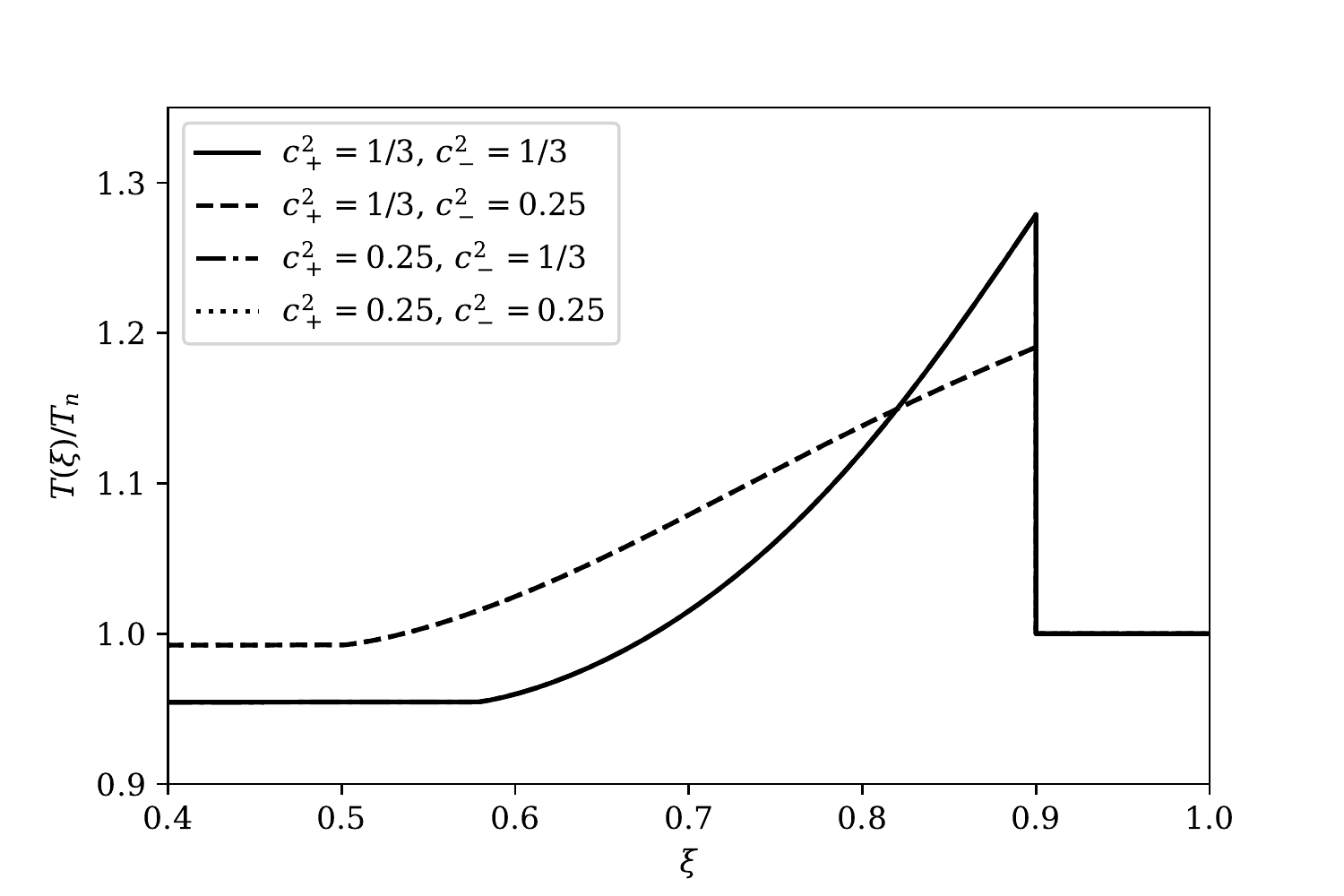}
		\end{minipage}
	}%
	\centering
	\caption{The temperature profile of the three hydrodynamical modes: the weak deflagration (upper left $v_w = 0.4$), hybrid (upper right $v_w = 0.6$), and weak detonation (bottom $v_w = 0.9$) with $\alpha_{\bar{\theta} n} = 0.3$.}\label{Tprofile}
\end{figure}
During a FOPT process, the liberated energy is not fully converted into the kinetic energy of the surrounding fluid.
Hence the kinetic energy fraction is smaller than $\mathcal{O}(1)$.
In fact the rest of the liberated energy reheats the fluid that surrounding the expanding bubble wall.
Based on the DSVM, we can derive the temperature profile with Eq.~\eqref{Tprof}.
Figure~\ref{Tprofile} depicts the temperature profile of three hydrodynamical modes which are weak deflagration, hybrid, and weak detonation respectively, and show reheating phenomenon of the expanding bubble wall.
The two upper plots of Fig.~\ref{Tprofile} represent the temperature profiles of deflagration ($v_w = 0.4$) and hybrid ($v_w = 0.6$) with different sound velocities of the symmetric and broken phase for a given strength parameter $\alpha_{\bar{\theta} n} = 0.3$,
and these plots show how the lower sound velocity of the broken phase could reduce the reheating effects in front of the bubble wall.
However, lower sound velocity of the symmetric phase could enhance the reheating effects.
The bottom panel of Fig.~\ref{Tprofile} shows the temperature profile of detonation mode with $\alpha_{\bar{\theta} n} = 0.3$,
and we find the reheating effect can only be altered by changing the sound velocity of the broken phase.
Decreasing the sound velocity of the broken phase weakens the reheating effect of detonation mode.
In Sec.~\ref{md}, we have calculated the sound velocity of the broken phase with nucleation and percolation temperature for simplicity.
A more precise approach to compute the sound velocity is  to use the temperature far behind the wall, which can be derived by solving the temperature profile.
However, Fig.~\ref{Tprofile} shows that the temperature far behind the wall is not significantly different from the nucleation temperature for detonation mode.
Therefore, we can use nucleation temperature or the percolation temperature as a proper approximation of the temperature of the broken phase.
For the deflagration mode, the reheating effect just in front of the wall may give some influences to the baryogenesis, which is studied in Ref.~\cite{Megevand:2000da}.

\subsection{A fully model-dependent analysis}
The key point of the analysis shown in Sec.~\ref{md} is to match a realistic model on a benchmark EOS (the DSVM of EOS is applied in this work); this can give a model-independent approach that simplifies the study of phase transition.
However, the accuracy of this method depends on whether the benchmark EOS can describe the phase transition process appropriately.
In a realistic phase transition process, the sound velocity should be temperature dependent.
This temperature-dependent behavior can be observed from Eqs.~\eqref{jb} and \eqref{jf}, which are the high-temperature expansion of thermal correction, and the definition of the sound velocity.
In the massless limit, namely $y=m/T=0$, the square of sound velocity is $1/3$. 
However, with the increasing of the mass, the deviation gradually increases.
Especially, when the mass becomes comparable to the phase transition temperature, the deviation becomes non-negligible. And from this aspect of thermal expansion terms, the daisy resummation might alleviate the sound velocity deviation. 
There exist two daisy resummation schemes, Arnold-Espinosa scheme \cite{Arnold:1992rz} and Parwani scheme \cite{Parwani:1991gq}, and they usually give different predictions.
Thus, the precise study of the resummation effects on the sound velocity deviation is left in our future work.  
Obviously, the strong and ultra supercooling cases could give a significant modification to the sound velocity.
The models with extra massive fermions might lead to large deviation as discussed in Ref.~\cite{Leitao:2014pda}.
The two-step phase transition and the standard model-like phase transition could also lead to sound velocity deviation~\cite{Giese:2020znk,Giese:2020rtr}.

Due to the reheating effect, the temperature is position dependent, hence the sound velocity is eventually position dependent,
and this should make the fluid equation more complicated, but we can use a fully numerical calculation to derive corresponding quantities. 
Here we exemplify this model-dependent analysis, and leave a further study to a future work.
For a given model, its free energy can be expressed as Eq.~\eqref{freeen}
and according to the free energy we can obtain the EOS
\begin{equation}
p = - V_T(\phi,T) - V_{T = 0}(\phi),\quad e = -T\frac{\partial V_T(\phi,T)}{\partial T} + V_T(\phi,T) + V_{T=0}(\phi,T)\,\,.
\end{equation}
Based on this EOS we can perform the same analysis in this work, and give the corresponding prediction of GW signals and SNR.

For the precise calculations of the GW spectra, there is another important parameter, the bubble wall velocity, which affects the results significantly. 
Since in most studies, the bubble wall velocity is taken as an input parameter,
and the GW spectra strongly depend on the bubble wall velocity, namely, the kinetic energy is strongly related to the bubble wall velocity.
In principle, this bubble wall velocity can be calculated by solving the equation of motion of the phase transition order-parameter fields and the collision terms for a given new physics model. It is also model dependent.
For this Higgs sextic model, our recent study \cite{Wang:2020zlf} calculates the bubble wall velocity beyond leading-log approximation.

\subsection{Multistep phase transition}
\begin{figure}[t]
	\centering
	\subfigure{
		\begin{minipage}[t]{1\linewidth}
			\centering
			\includegraphics[scale=0.35]{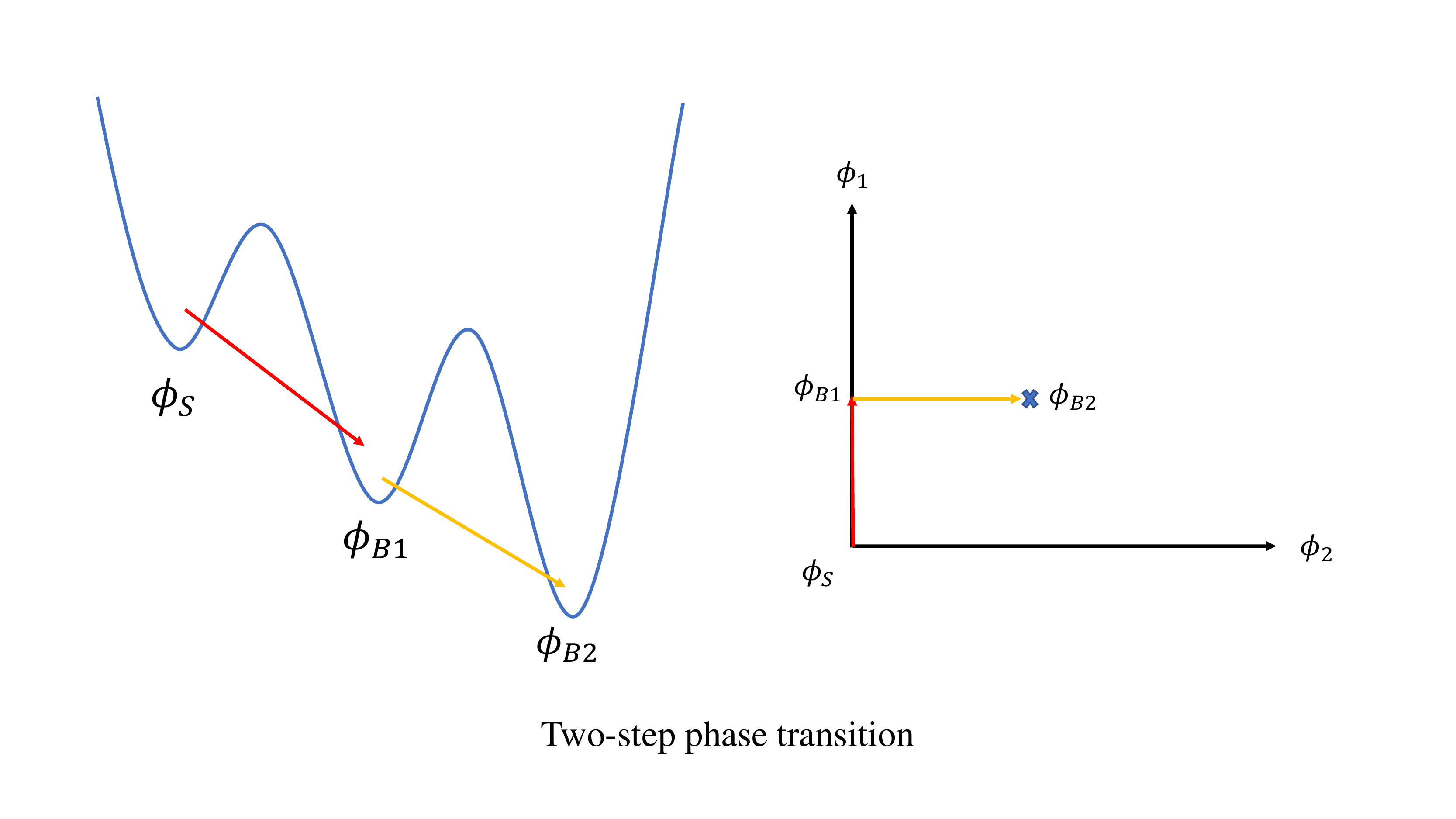}
		\end{minipage}%
	}%
	\centering
	\caption{Illustration of the two-step phase transition process.}\label{tpt}
\end{figure}
For a given new physics model with $N$ scalar fields ($N>1$), each scalar field could obtain a VEV
and there could occur multi-step FOPTs in the $N$ scalar space.
Here we take a two-step phase transition as an example, the first step phase transition is the same as the one-step phase transition.
Illustration of the two-step phase transition process is shown in Fig.~\ref{tpt}.
However, the second phase transition shows some differences. For example, the sound velocity in $\phi_{B1}$ phase deviates from $1/\sqrt{3}$. 
Specifically, for the phase transitions with bubbles nucleated inside the bubbles~\cite{Vieu:2018zze,Croon:2018new},
the effects of different sound velocity may be even more important. In this case, the EOS and sound velocities in each phase may have significant differences.

\section{Conclusion}\label{cn}
Taking the Higgs sextic effective model as a representative model for generic new physics models,
we have calculated the kinetic energy fraction by solving the hydrodynamics equations and the fluid profile beyond the bag model approximation for the two phases of the plasma.
For strong supercooling and ultra supercooling cases, the deviation of the sound velocity in the broken phase from $1/\sqrt{3}$ is significant. In these two cases, the
nucleation temperature or percolation temperature is much lower than the masses of the Higgs and gauge bosons. This would lead to deviation from the pure radiation phase.
Based on the different sound velocities model, which is more realistic than the bag model, different sound velocities could occur in symmetric and broken phase.
Choosing proper phase transition parameters, characteristic temperature and more realistic model of EOS, we can give more precise and reliable predictions of the GW signal.
The sound velocity of broken phase can significantly affect the detectability of the GW signal.  
The approach shown in this work could help the future GW  experiments to unravel the underlying physics by matching the precise spectra  prediction to the data, and could be directly used in other new physics models with a strong FOPT.

\begin{acknowledgments}
F.P.H. thanks Yi-Ming Hu and Zheng-Cheng Liang for helpful discussions and  providing  the updated sensitivity of TianQin project.
F.P.H. is supported in part by the McDonnell Center for the Space Sciences and Guangdong Major Project of Basic and Applied Basic Research (Grant No. 2019B030302001).
X.W. and X.M.Z. are supported in part by  the Ministry of Science and Technology of China (2016YFE0104700), the National Natural Science Foundation of China (Grant No. 11653001), the CAS pilot B project (XDB23020000).
\end{acknowledgments}

\noindent
\textbf{Note added}: While this paper was under completion, we noticed Ref.~\cite{Giese:2020znk} appeared on arXiv, partially overlapping with this work.

\newpage

\begin{appendices}

\section{Fluid profile}
\label{profile}

\begin{figure}[t]
	\centering
	
	\subfigure{
		\begin{minipage}[t]{0.5\linewidth}
			\centering
			\includegraphics[scale=0.5]{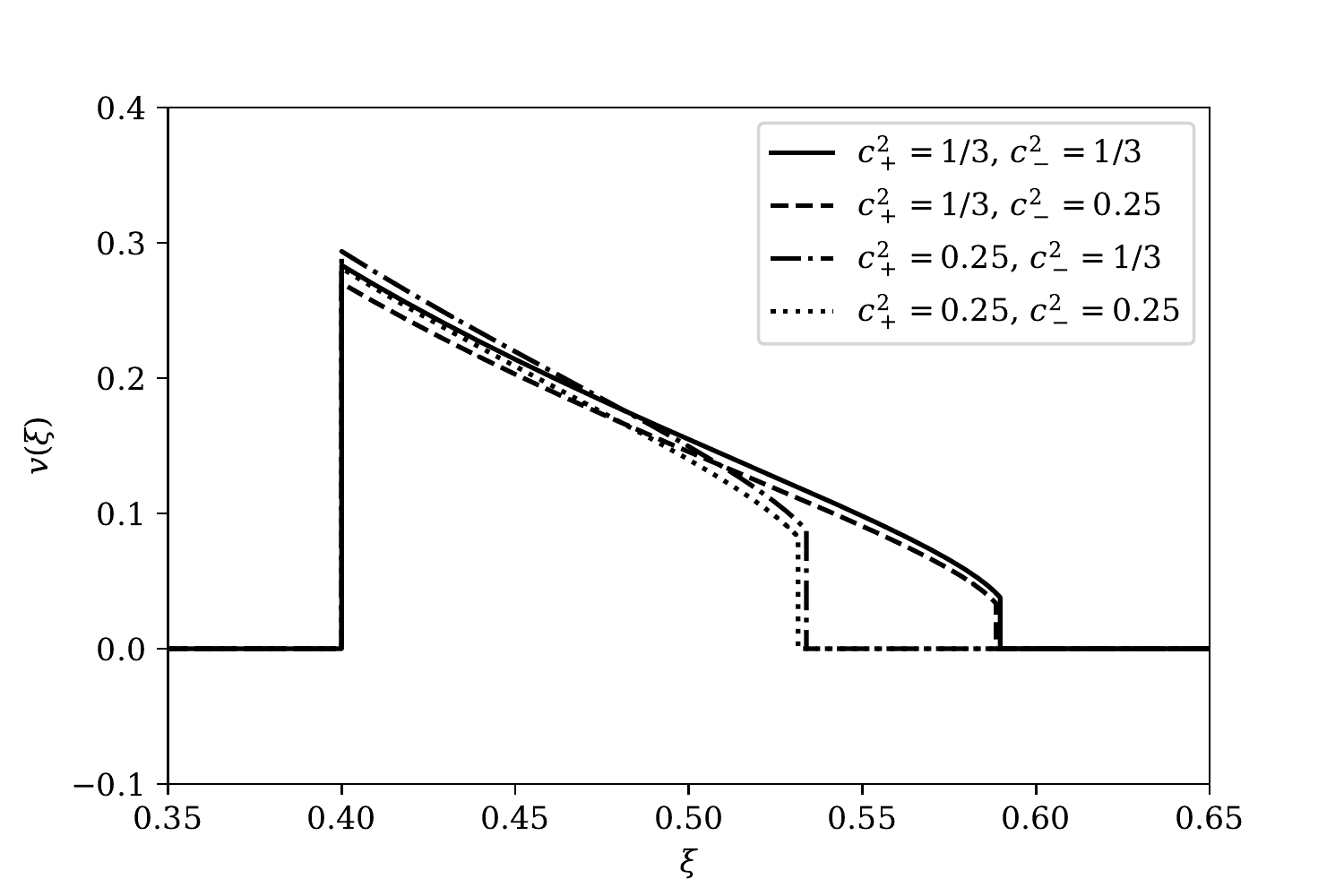}
		\end{minipage}%
	}%
	\subfigure{
		\begin{minipage}[t]{0.5\linewidth}
			\centering
			\includegraphics[scale=0.5]{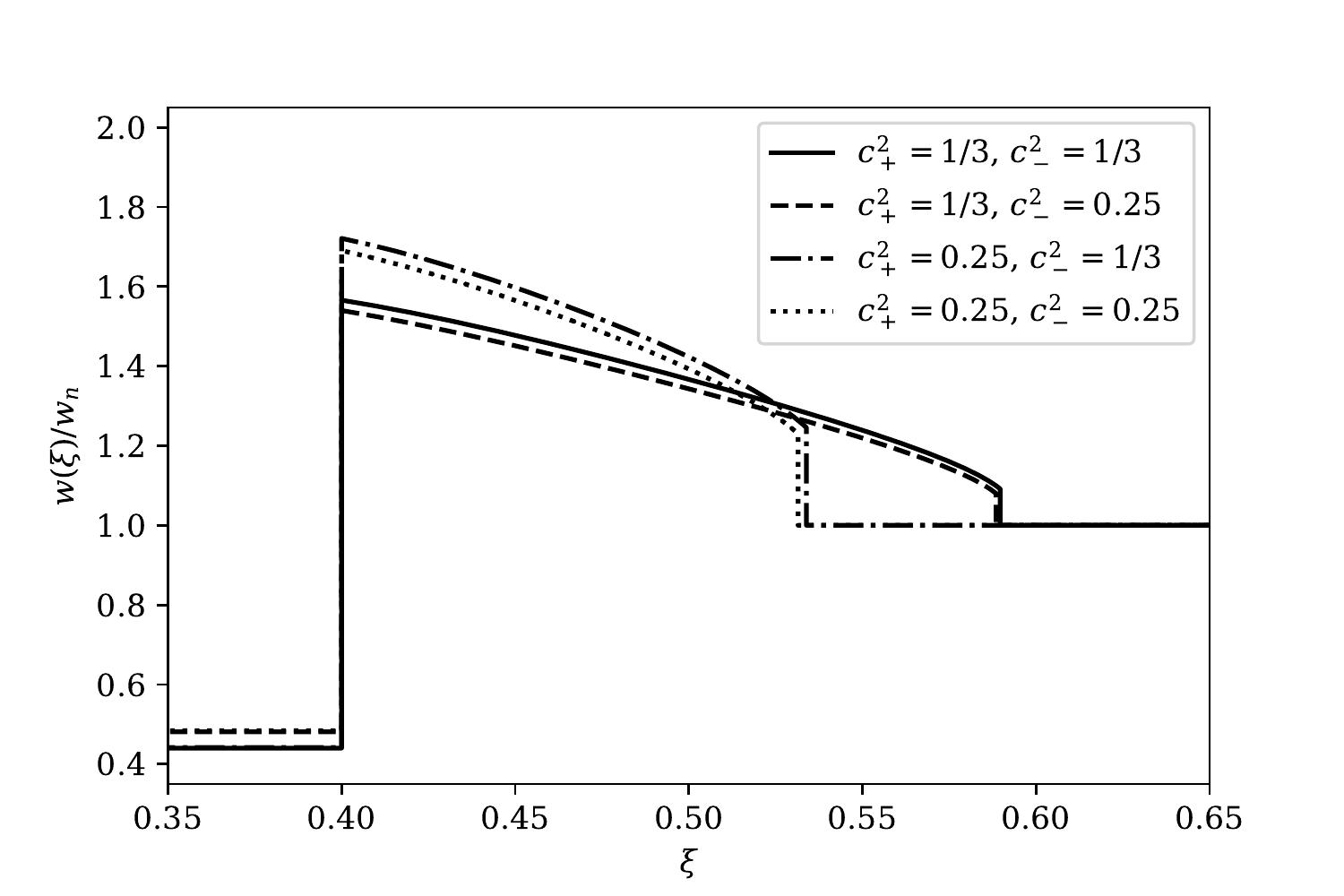}
		\end{minipage}%
	}%
	\quad
	\subfigure{
		\begin{minipage}[t]{0.5\linewidth}
			\centering
			\includegraphics[scale=0.5]{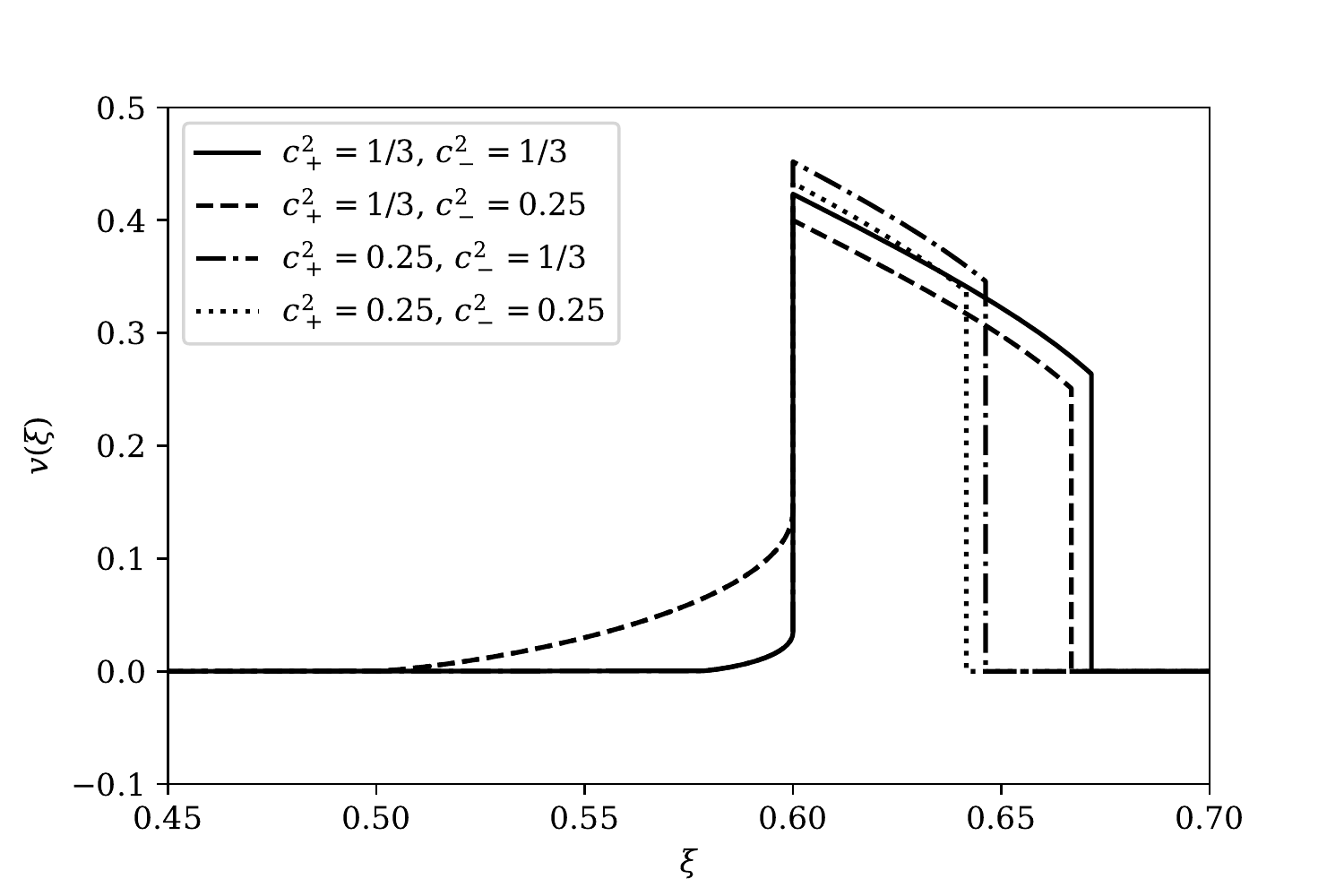}
		\end{minipage}
	}%
	\subfigure{
		\begin{minipage}[t]{0.5\linewidth}
			\centering
			\includegraphics[scale=0.5]{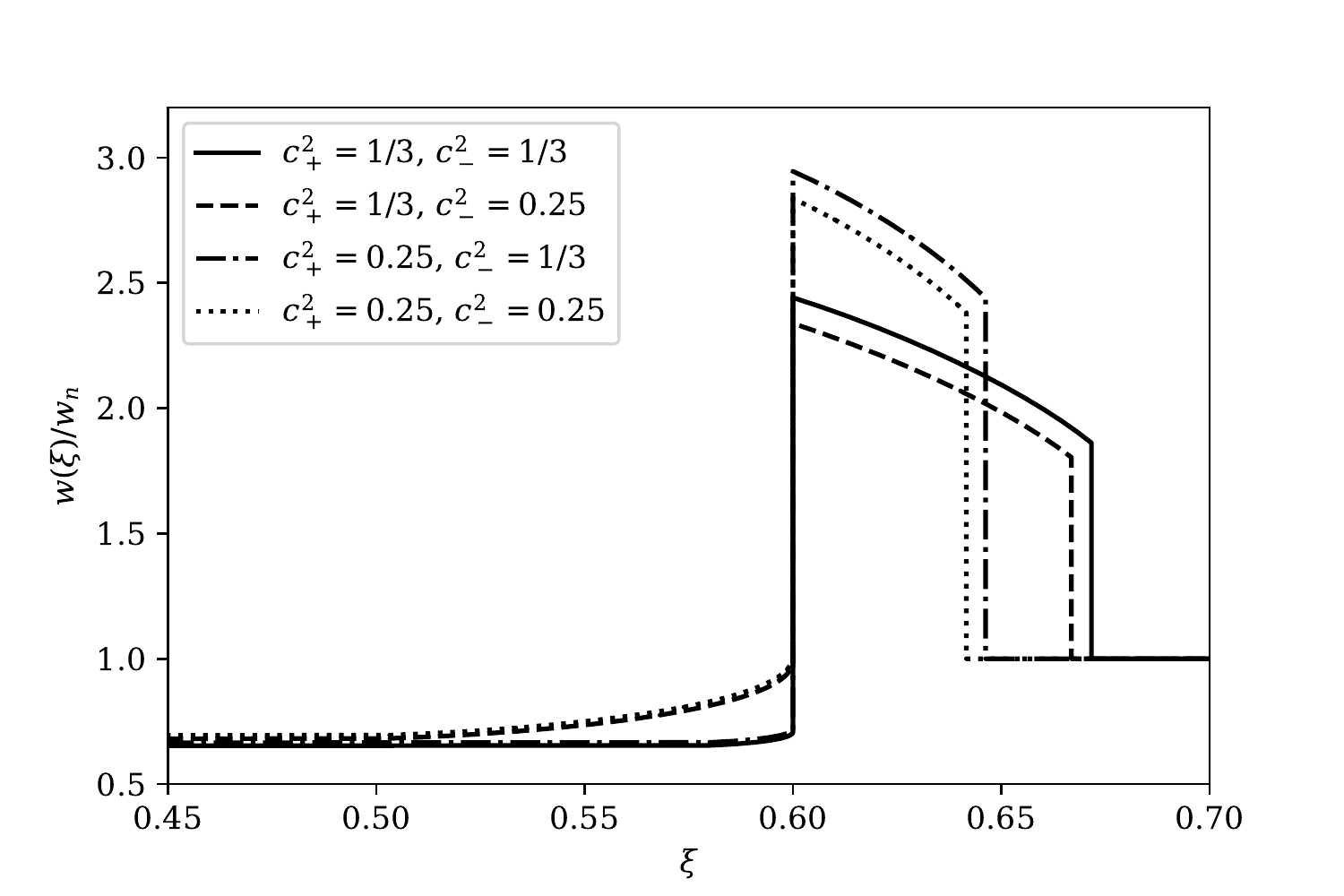}
		\end{minipage}
	}%
	\quad
	\subfigure{
		\begin{minipage}[t]{0.5\linewidth}
			\centering
			\includegraphics[scale=0.5]{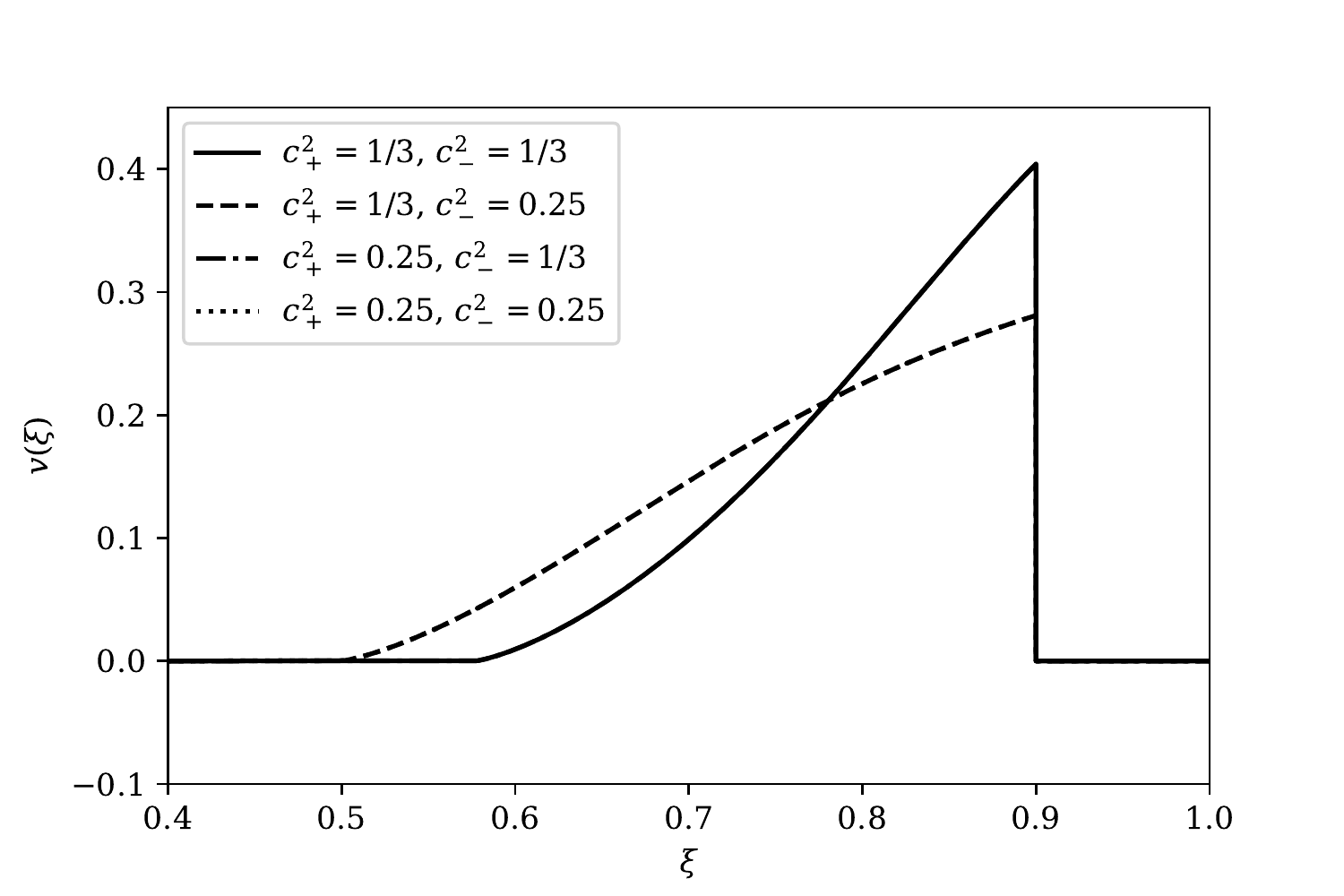}
		\end{minipage}
	}%
	\subfigure{
		\begin{minipage}[t]{0.5\linewidth}
			\centering
			\includegraphics[scale=0.5]{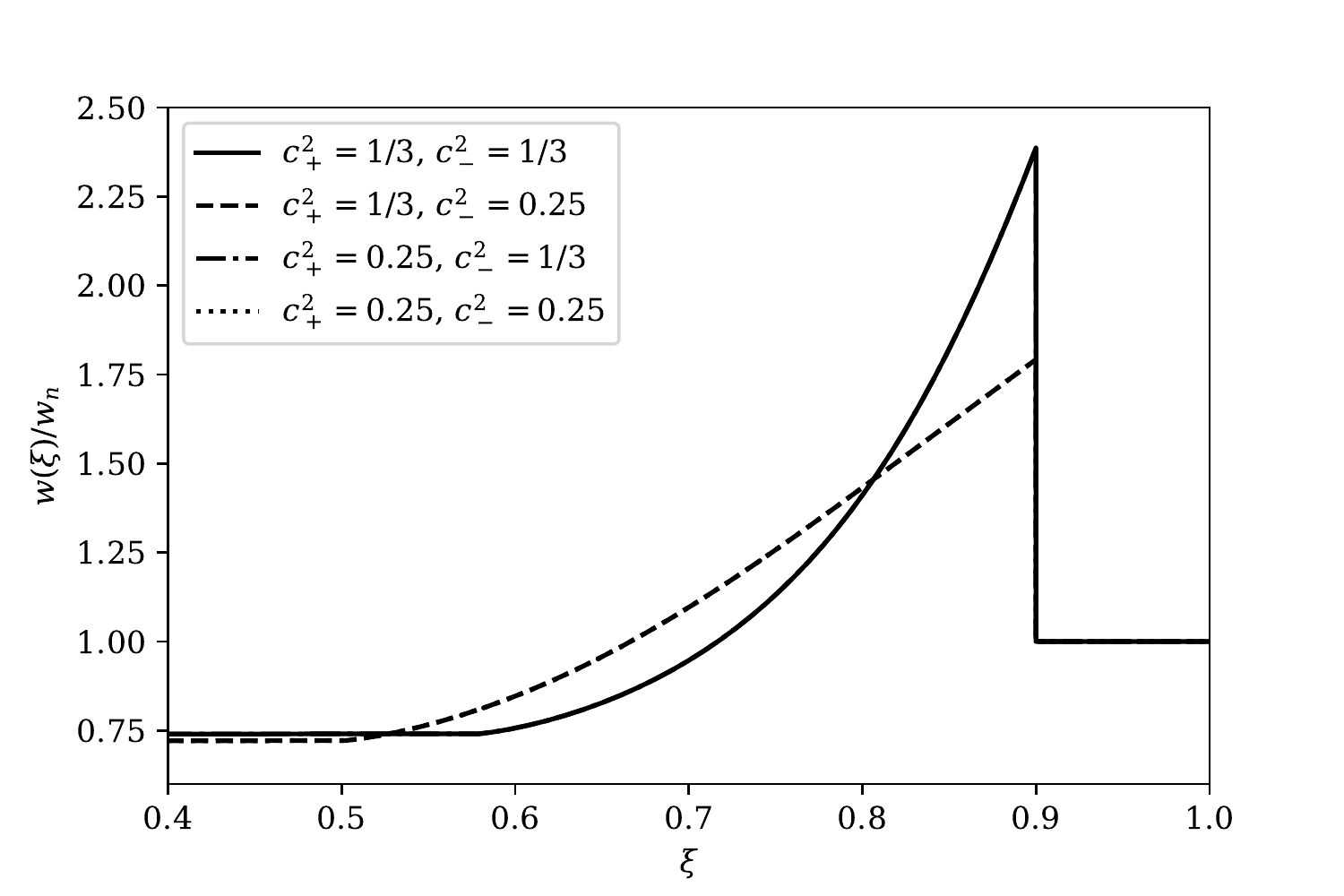}
		\end{minipage}
	}%
	\centering
	\caption{Velocity and enthalpy profiles for the weak deflagration (top $v_w = 0.4$), hybrid (middle $v_w = 0.6$) and weak detonation (bottom $v_w = 0.9$) with $\alpha_{\bar{\theta} n} = 0.3.$}\label{vepd}
\end{figure}

As shown in Fig.~\ref{vpvm}, the hydrodynamical processes can be roughly divided into two kinds, which are detonation ($v_-< v_+$) and deflagration ($v_- > v_+$),
and we can further divide these two kinds into six modes, which are weak detonation, strong detonation, Jouguet detonation, weak deflagration, strong deflagration, and hybrid.
However, the stability analysis \cite{Huet:1992ex,Megevand:2013yua,Megevand:2014yua} shows only three kinds of modes can be realized in a FOPT process, namely, weak detonation, weak deflagration, and hybrid.
In the following, we describe these modes and show the velocity and enthalpy profiles with $\alpha_{\bar{\theta} n} = 0.3$ in Fig.~\ref{vepd}.

\subsection*{Detonation}
For the bubble wall moves with $v_w > v_J^{\rm det}(\alpha_{\bar{\theta} n})$, we have a weak detonation mode.
In this the case, the fluid is unperturbed in front of the bubble wall with respect to the reference frame of the plasma, we have $\tilde{v}_+ = 0$ ($v$ and $\tilde{v}$ represent the fluid velocity in the wall frame and the plasma frame respectively).
Hence we have $v_w = v_+$, and the fluid velocity behind the wall must have a velocity $\tilde{v}_- > 0$.
Therefore, the boundary conditions to solve the profile of detonation mode can be obtained as 
\begin{equation}
\tilde{v}_+ = 0, \quad v_+ = v_w, \quad v_- = v_-(\alpha_{\bar{\theta}+}, v_+), \quad v(v_w) = \tilde{v}_- = \mu(v_w, v_-)\,\,,
\end{equation}
and here we have $\alpha_{\bar{\theta} n} = \alpha_{\bar{\theta}+}$.
Using the boundary conditions for the enthalpy profile $w_+ = w_n$ and the velocity $v_-$, $v_+$ derived above, based on the second equations of the matching condition, we can obtain the enthalpy just behind the bubble wall as 
\begin{equation}
w_- = w_+\left(\frac{v_w}{1 - v_w^2}\right)\left(\frac{1-v_-^2}{v_-}\right)\,\,,
\end{equation} 
and from Eqs.~\eqref{constantmodel}, we can derive
\begin{equation}
w_+ = (1 + c_+^2)a_+T_+^4\,\,.
\end{equation}
Since $T_+ = T_n$ for the detonation mode, we need use the following relations to obtain $T_-$
\begin{equation}
\frac{w_n}{w_-} = \frac{a_+T_+^{4}}{a_-T_-^{4}}\,\,.
\end{equation}
To determine $T_-$, $a_+$ and $a_-$ should be fixed.
Based on the Higgs sextic effective model we assume the particle content in the symmetric phase is the standard model particles and the top quark and Higgs boson decouple from the plasma in the broken phase.
Hence we have $a_+ = 106.75\pi^2/30$ and $a_- = 95.25\pi^2/30$.
The velocity and enthalpy profile of detonation mode with different sound velocities are shown in bottom panel of Fig.~\ref{vepd}.

\subsection*{Deflagration}
\begin{figure}[t]
	\centering
	
	\subfigure{
		\begin{minipage}[t]{1\linewidth}
			\centering
			\includegraphics[scale=0.45]{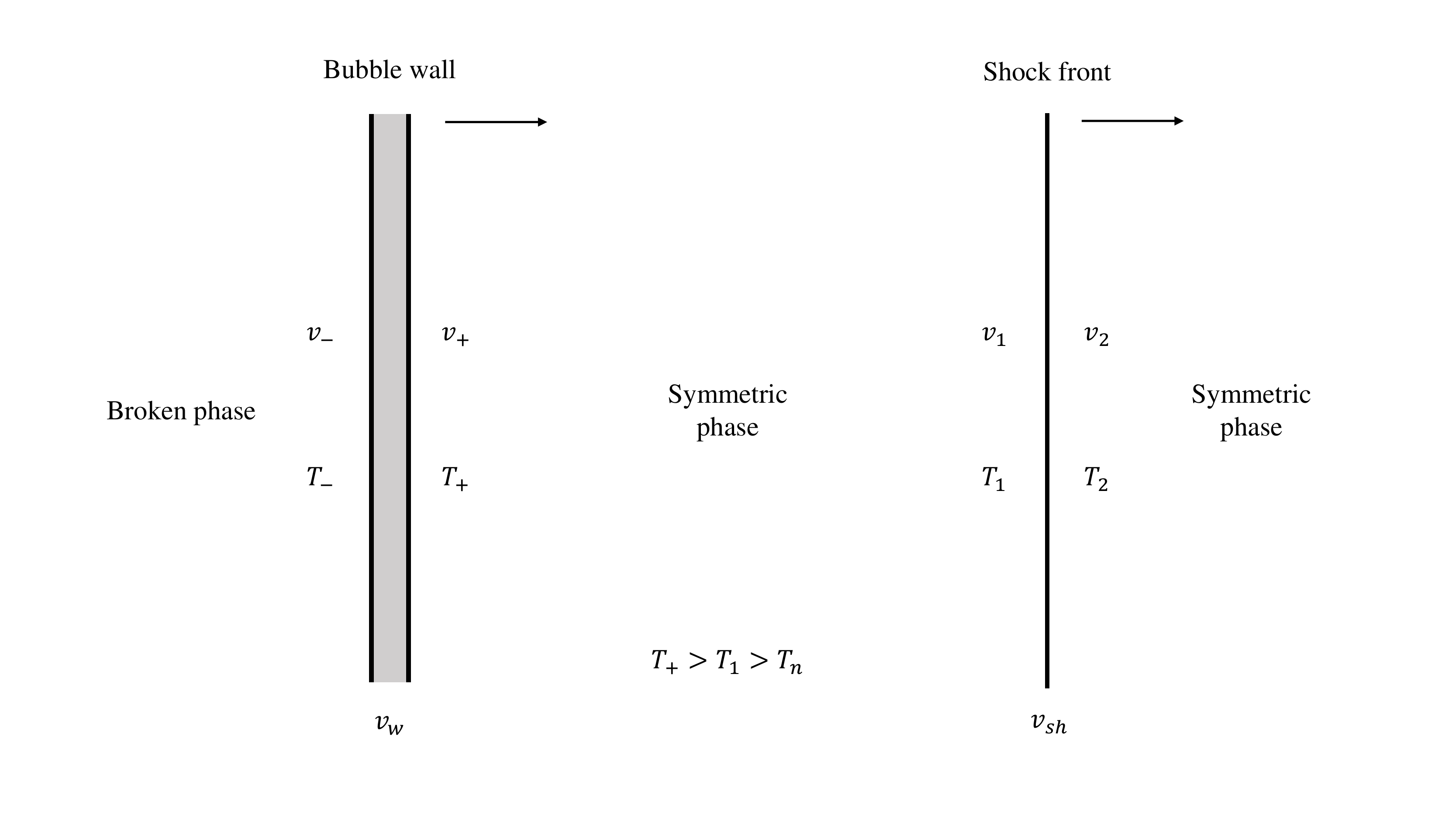}
		\end{minipage}%
	}%
	\centering
	\caption{An illustration of the deflagration mode. For the detonation mode, the shock front should disappear.}\label{def}
\end{figure}

If the bubble wall is subsonic with respect to the reference frame of plasma, $v_w < c_-$, we have a deflagration mode.
As shown in Fig.~\ref{def}, the deflagration mode should form a shock front which is in front of the bubble wall.
Here $v_1$ and $v_2$ is the fluid velocity behind and in front of the shock front with respect to the shock front frame.
The index 1 is for quantities behind the shock and the index 2 is for quantities in front of the shock.
From Eqs.~\eqref{constantmodel}, we have the following relations
\begin{equation}
v_1v_2 = \frac{p_2 - p_1}{e_2 - e_1},\quad\frac{v_1}{v_2} = \frac{e_2 + p_1}{e_1 + p_2}\,\,.
\end{equation}
Since the shock wave is in the symmetric phase, the EOS is the same on the both side of shock front, we have
\begin{equation}
v_1v_2 = c_+^2,\quad\frac{v_1}{v_2} = \frac{T_2^4 + c_+^2T_1^4}{T_1^4 + c_+^2T_2^4}\,\,,\label{shock}
\end{equation}
and for the shock front it has been proven that $v_1 < c_+ < v_2$ \cite{Espinosa:2010hh,Leitao:2010yw,Leitao:2014pda}.
Hence in the plasma frame, we have $\tilde{v}_1 = \mu(v_{sh},v_1) > \tilde{v}_2 = \mu(v_{sh},v_2)$.
Since in the plasma frame the fluid velocity should vanish in front of the shock, we have $v_2 = v_{sh}$.
Hence, the following relation can be derived from Eqs.~\eqref{shock}:
\begin{equation}
v_{sh} = \frac{1 - c_+^2}{2}\tilde{v}_1 + \sqrt{\left(\frac{1 - c_+^2}{2}\tilde{v}_1\right) + c_+^2}\,\,.
\end{equation}
This condition determines the position of the shock front, which occurs before the singular point $\mu(\xi, v) = c_+$ is reached, and show that $v_{sh} > c_+$.
From the second equation of Eqs.~\eqref{shock}, we can derive
\begin{equation}
\left(\frac{T_2}{T_1}\right)^{4} = \frac{w_2}{w_1} = \frac{c_+^2(1 - v_{sh}^2)}{v_{sh}^2 - c_+^4}\,\,,\label{srelat}
\end{equation}
where $T_2 = T_n$, $w_n = w_2$. 

As shown in the above, the bubble wall of the deflagration mode is preceded with a shock front, hence behind the wall the fluid velocity should also vanish with respect to the plasma frame.
Therefore, we have $\tilde{v}_- = 0$ and $v_+ < v_- = v_w < c_-$.
We can choose the fluid velocity $\tilde{v}_+$ (just in front of the bubble wall) and $\tilde{v}_1$ (behind the shock front) as the initial condition for the fluid equation.
Here we choose $\tilde{v}_+$ as the initial condition, and it can be derived form the following relations,
\begin{equation}
\tilde{v}_- = 0, \quad v_- =  v_w, \quad v_+ = v_+(\alpha_{\bar{\theta}+}, v_-), \quad \tilde{v}_+ = \mu(v_w, v_+)\,\,.
\end{equation}
Note that $\alpha_{\bar{\theta}+}\ne\alpha_{\bar{\theta} n}$ for deflagration mode, since the shock wave reheats the plasma in front of the bubble wall,
and for the DSVM of EOS, we have
\begin{equation}
\alpha_{\bar{\theta}+} = \frac{1}{3}\left[\frac{1 - c_+^2/c_-^2}{1 + c_+^2} + \frac{(1 + 1/c_-^2)\Delta\epsilon}{w_+}\right]\,\,.
\end{equation}
Hence we can firstly find $v_{sh}$ and $\alpha_{\bar{\theta}+}$ with 
\begin{equation}
\frac{\alpha_{\bar{\theta}+} - b}{\alpha_{\bar{\theta} n} - b} = \frac{w_n}{w_+},\quad b=\frac{1}{3}\left[\frac{1 - c_+^2/c_-^2}{1 + c_+^2}\right]\,\,,
\end{equation}
and $w_+$ can be given by Eq.~\eqref{Tprof},
\begin{equation}
\frac{w_1}{w_+} = \exp\left[\int_{\tilde{v}_+}^{\tilde{v}_1}\left(1 + \frac{1}{c_+^2}\right)\gamma^2\mu dv\right]\,\,,
\end{equation} 
where $w_1$ can be obtain by Eq.~\eqref{srelat}.
The velocity and enthalpy profile of deflagration mode are shown in top panel of Fig.~\ref{vepd}.

\subsection*{Hybrid}
For the different sound velocity model, the hybrid can be further divided into two modes, which are supersonic deflagration and subsonic detonation.
The subsonic detonation is only possible for the model with $c_- < c_+$,
and both modes should fulfill that the bubble wall velocity $v_w$ is higher than both $v_+$ and $v_-$.
The velocity and enthalpy profile of hybrid mode are shown in middle panel of Fig.~\ref{vepd}.
\subsubsection{\textbf{Supersonic deflagration}}
Usually, if $c_- < v_w < v_J^{\rm det}(\alpha_{\bar{\theta} n})$ the fluid propagates with a supersonic deflagration mode.
It should fill the gap between $c_-$ and $v_J^{\rm det}(\alpha_{\bar{\theta} n})$.
The supersonic deflagration mode can be treated as a superposition of detonation and deflagration, provided the bubble wall is supersonic with respect to the broken phase.
The entropy consideration, which combines with the hydrodynamic constrains for deflagration and detonation mode, enforces $v_- = c_-$.
Hence the rarefaction wave behind the bubble wall has to be a Jouguet type.
We can derive the following relation to solve Eq.~\eqref{vprofile}
\begin{equation}
v_- = c_-,\quad\tilde{v}_- = \mu(v_w, v_-),\quad v_+ = v_J^{\rm def}(\alpha_{\bar{\theta}+}),\quad\tilde{v}_+ = \mu(v_w, v_+)\,\,.
\end{equation}
In our numerical strategy, we first perform the calculation of the deflagration part to give $\alpha_{\bar{\theta}+}$ and corresponding profiles, then calculate the profile of detonation part as mentioned above.

\subsubsection{\textbf{Subsonic detonation}}
For the sound velocity of the broken phase $c_-$ is smaller than the sound velocity of the symmetric phase $c_+$, we may observe a hydrodynamical mode with $\tilde{v}_- > \tilde{v}_+$ (ie. $v_+ > v_-$),
and $v_+ < c_+$, hence in the reference frame of the plasma, the fluid profile propagates with a subsonic mode.
Since the constraints of deflagration and detonation and the entropy consideration still apply, we also have the following relation\ $v_- = c_-$.
Therefore the rarefaction wave still propagates with a Jouguet type.
From right panel of Fig.~\ref{vpvm}, we can observe that only when $\alpha_{\bar{\theta}+}$ is small enough this hydrodynamical mode can occur, and the stability \cite{Megevand:2014yua,Megevand:2013yua,Huet:1992ex} of this mode is still controversial; hence we do not consider this possibility in this study.

\section{Efficiency parameter}
\begin{figure}
	\centering
	
	\subfigure{
		\begin{minipage}[t]{1\linewidth}
			\centering
			\includegraphics[scale=0.75]{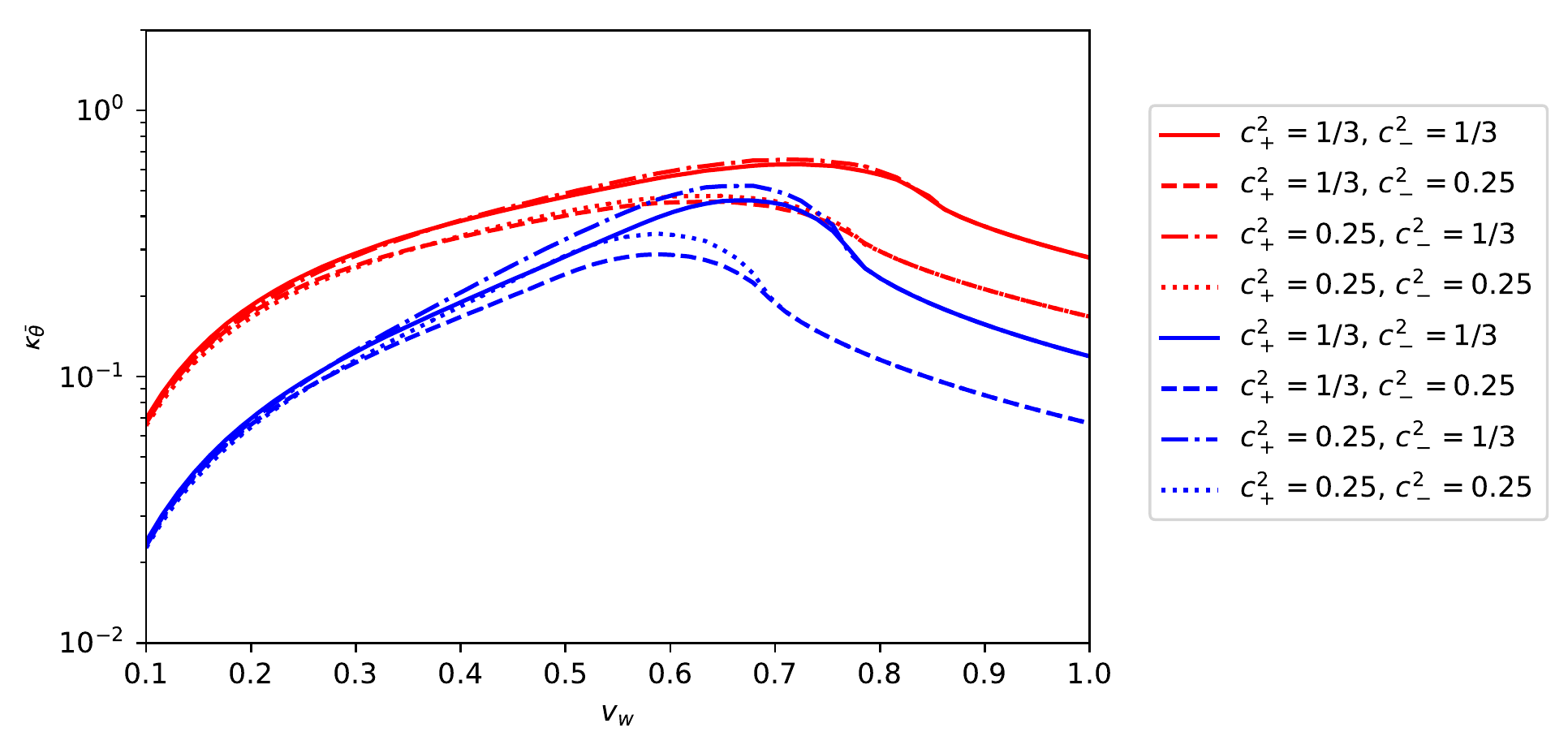}
		\end{minipage}%
	}%
	\centering
	\caption{Illustration of the efficiency parameter dependence on bubble wall velocity for different phase transition strength and sound velocities. The red lines and the blues lines correspond to $\alpha_{\bar{\theta}n} = 0.3$ and $\alpha_{\bar{\theta} n} = 0.1$ respectively. The solid lines represent $c_+^2 = c_-^2 = 1/3$, the dashed lines show $c_+^2 = 1/3$ and $c_-^2 = 0.25$, the dash-dotted lines denote $c_+^2 = 0.25$ and $c_-^2 = 1/3$, the dotted lines depict $c_+^2 = 0.25$ and $c_-^2 = 0.25$.}\label{kappa}
\end{figure}

In Fig.~\ref{kappa}, we show the efficiency parameter dependence on bubble wall velocity for different phase transition strength and sound velocities.
The red lines and the blues lines correspond to $\alpha_{\bar{\theta}n} = 0.3$ and $\alpha_{\bar{\theta} n} = 0.1$ respectively. 
The solid lines represent $c_+^2 = c_-^2 = 1/3$ (i.e. the bag model case), the dashed lines show $c_+^2 = 1/3$ and $c_-^2 = 0.25$, the dash-dotted lines denote $c_+^2 = 0.25$ and $c_-^2 = 1/3$, the dotted lines depict $c_+^2 = 0.25$ and $c_-^2 = 0.25$.
As shown in this figure, the lower sound velocity of the broken phase gives smaller efficiency parameter.
However, the lower sound velocity of the symmetric phase induces larger efficiency parameter. 
The differences of the efficiency parameter between the bag model and the DSVM is smaller for the deflagration mode,
while for the hybrid mode and detonation mode, the corresponding differences become much larger.
Since detonation mode can generally trigger a relatively strong gravitational wave signal, it is crucial to give a correct EOS (i.e. proper sound velocity of both phase) to describe the phase transition process.
Then we can get more precise prediction for the GW.

\end{appendices}

\end{document}